\documentclass[11pt, a4paper]{article}

\usepackage{color}
\usepackage{graphicx}
\usepackage{amssymb}
\usepackage{amsmath}
\usepackage[english]{babel}
\usepackage[T1]{fontenc}
\usepackage[noadjust]{cite}
\usepackage{hyperref} 

\usepackage[utf8]{inputenc}
\usepackage{authblk}

\newcommand\eeq{\end{equation}}
\newcommand\beq{\begin{equation}}
\newcommand\eea{\end{eqnarray}}
\newcommand\bea{\begin{eqnarray}}

\begin{document}

\linespread{1.1}

\title{{\bf Electroweak metastability and Higgs inflation}}

\author[1,2]{ {\large\sc Isabella Masina} \thanks{masina@fe.infn.it}}
\author[3]{ {\large\sc Mariano Quiros} \thanks{quiros@ifae.es}}

\affil[1]{\small\it Dept.\,of Physics and Earth Science, Ferrara University, Via Saragat 1, 44122 Ferrara, Italy }
\affil[2]{\small\it Istituto Nazionale di Fisica Nucleare (INFN), Sez.\,di Ferrara, Via Saragat 1, 44122 Ferrara, Italy }
\affil[3]{\small\it Institut de F\'{\i}sica d'Altes Energies (IFAE) and The Barcelona Institute of  Science and Technology (BIST), Campus UAB, 08193 Bellaterra (Barcelona) Spain}

\date{}

\maketitle

\begin{abstract}

Extrapolating the Standard Model Higgs potential at high energies, we study the barrier between the electroweak and Planck scale minima.
The barrier arises by taking the central values of the relevant experimental inputs, that is the strong coupling constant and the top quark and Higgs masses. 
We then extend the Standard Model by including a non-minimal coupling to gravity, and explore the phenomenology of the Higgs inflation model. 
We point out that even configurations that would be metastable in the Standard Model, 
become viable for inflation if the non-minimal coupling is large enough to flatten the Higgs potential at field values below the barrier; 
we find that the required value of the non-minimal coupling is smaller than the one needed for the conventional Higgs inflation scenario
(which relies on a stable Standard Model Higgs potential, without any barrier); 
in addition, values of the top mass which are larger than those required in the conventional scenario are allowed.

 \end{abstract}

\linespread{1.2}

\vskip 1.cm
\section{Introduction}

In the pure Standard Model (SM), and depending on the value of the electroweak parameters, there is the possibility of having a second minimum at high energies~\cite{Hung:1979dn, Cabibbo:1979ay, Froggatt:1995rt}.
As it is well known, the experimental data suggest that the electroweak vacuum is likely to be metastable rather than stable~\cite{Degrassi, Masina:2012tz, Buttazzo:2013uya, Bednyakov:2015sca}, 
although it will be difficult to exclude stability in future~\cite{Franceschini:2022veh, Hiller:2024zjp}. 
The study of the shape of the Higgs effective potential at high energy is a relevant issue, also in view of the possible role of the Higgs field as the inflaton;
for this sake, a region where the Higgs potential becomes sufficiently flat, for large enough values of the Higgs field, to meet the slow-roll conditions would be required.

In this paper, we will first reconsider the analysis of stability and metastability of the SM electroweak vacuum, in view of the most recent experimental results and theoretical developments; and secondly, we will do it for the case where the Higgs is non-minimally coupled to gravity. 
For the central values of the relevant experimental inputs, which are the strong coupling constant with five flavors, $\alpha_s^{(5)}$, 
the top quark pole mass, $m_t$, and the Higgs mass, $m_H$, in the SM potential a barrier arises between the electroweak minimum and a second high scale minimum. 
We study in detail the phenomenology of such a barrier, as its height and position.  
Fixing $\alpha_s^{(5)}$ and $m_H$ to their central values, there is a particular value of the top mass, $m_t^i$, 
corresponding to a stable inflection point configuration: for any $m_t > m_t^i$ the barrier thus takes place~\cite{Iacobellis:2016eof}. 
In particular, the electroweak and Planck scale vacua are degenerate for the critical value $m_t^c$~\cite{Froggatt:1995rt}.

The possibility of having an inflection point, or a shallow vacuum configuration, attracted interest in view of cosmological applications:
rolling down~\cite{Hamada:2014iga, Masina:2014yga, Bezcritical, Hamada:2014wna, Ezquiaga:2017fvi, Salvio:2017oyf}, or being trapped into such configurations~\cite{MasinaHiggsmass, Masinatop, Masinahybrid, Masina:2014yga}, 
the Higgs field might in principle induce a stage of primordial inflation.
It was shown that this scenario cannot however account for the observed cosmological parameters~\cite{Masina:2018ejw} 
(see also~\cite{Isidori:2007vm,Gialamas:2022gxv,Gialamas:2023emn} for previous related ideas). 
For SM stable configurations with the barrier, \textit{i.e.}~for $m_t^c > m_t > m^t_i$, and minimal coupling to gravity, we find that the barrier at its maximum turns out to be larger than the upper limit on the 
inflaton potential coming from CMB data. 
Then, configurations as an inflection point, or a shallow minimum, are not viable for inflation~\cite{Masina:2018ejw}, unless modifying the potential \cite{Salvio:2018rv}.

Extensions of the SM are thus needed.
By introducing a non-minimal coupling to gravity, $\xi$, that flattens\footnote{For an alternative approach postulating flatness, but without the non-minimal coupling, 
see\,\cite{Hamada:2013mya}.} the Higgs potential at field values larger than about $M_P/\sqrt{\xi}$,
where $M_P=1/\sqrt{8\pi G_N}$ is the reduced Planck scale, the Higgs might successfully play the role of the inflaton~\cite{BezHiggs}.
We thus secondly introduce the non-minimal coupling to gravity, $\xi |\mathcal H|^2 R$, and study in detail how the Higgs potential is modified. 
We do this for stable configurations without any barrier ($m_t<m^i_t$), as traditionally done in the previous literature~\cite{BezHiggs, Barvinsky, DeSimone:2008ei, Beztwoloop, Barvinskyrenorm, Bezrukov:2010jz, Bezrukov:2012sa, Allison, Bezrukov:2014ina, Bezrukov:2017dyv}, as well as for configurations with the barrier.
The latter are both the stable configurations where the second minimum is higher than the electroweak one (such that $m_t^c>m_t>m^i_t$),
and the metastable configurations where the second minimum is much deeper than the electroweak one ($m_t > m_t^c$).

Standard Model stable configurations without barrier require low values of the top pole mass, $m_t$.
According to our analysis, they correspond to the lower $2\sigma$ range of the PDG 2022 world average $m_t=172.5$ GeV~\cite{PDG2022}. 
In this case, as it is well known, large values of $\xi\gg 1$ are required in order to fit the cosmological observables derived from the cosmic microwave background
(CMB) data~\cite{BezHiggs, Barvinsky, DeSimone:2008ei, Beztwoloop, Barvinskyrenorm, Bezrukov:2010jz, Bezrukov:2012sa, Allison, Bezrukov:2014ina, Bezrukov:2017dyv}. 
We revisit this calculation and study in detail the dependence of $\xi$ on the top mass.

For SM metastable configurations, \textit{i.e.}~for $m_t>m_t^c$, we find that, for a suitable range of values of the non-minimal coupling $\xi$, 
it is possible to avoid the second high energy minimum of the Higgs potential: this happens when the potential is flattened before the barrier, so that it stays frozen to positive values in the region where the high energy minimum would take place for a minimal coupling to gravity.
We then focus on Higgs inflation in the case of metastable configurations, a scenario that, to the best of our knowledge, has not been previously studied, apart from Ref. \cite{Bezrukov:2014ipa} where threshold effects are considered\,\footnote{As first introduced in \cite{Bezrukov:2010jz} and discussed in detail in \cite{Bezrukov:2014ipa}, 
threshold effects at the transition scale $M_P/\xi$ encode the uncertainties associated with the specific UV completion of the SM non-minimally coupled to gravity and may allow for inflation even if the Higgs field is completely unstable.}.
In particular, we find that the values of $\xi$ which are required for successful inflation are even slightly smaller than those required in the well-studied case of stable configurations without barriers. As a result, values of the top mass which are slightly larger than those required in the conventional Higgs inflation scenario are allowed.

The paper is organized as follows.
In Sec.~\ref{sec:running} we study the running of the SM parameters and review the derivation of the Higgs effective potential. 
In Sec.~\ref{sec:barrier} we discuss the phenomenology of the barrier between the electroweak and Planck scale minimum.
In Sec.~\ref{sec:potential} we review the SM extension with a non-minimal coupling to gravity, including radiative corrections. 
In Sec.~\ref{sec:flattening} the values of $\xi$ required to avoid metastability are discussed.
In Secs.~\ref{sec:metric} and \ref{sec:Palatini} we discuss the phenomenology of Higgs inflation, for both stable and metastable configurations, in the metric and Palatini formalisms, respectively.
In Sec.~\ref{sec:unitarity} some comments about the unitarity issue in the model with non-minimal coupling to gravity are done.
In Sec.~\ref{sec:conclusions} we draw our conclusions.

\vskip 1.cm
\section{The running SM parameters}
\label{sec:running}

In this section we review the steps required to extrapolate the SM couplings at high energies.
The physical Higgs boson field $\phi$ belongs to the complex scalar Higgs doublet, $\mathcal{H}=e^{i\vec\chi\cdot\vec\sigma}(0,(\phi+v)/\sqrt{2})^T$, 
where $\vec\chi$ are the Goldstone bosons and $v$ is the electroweak scale. 
The tree-level Higgs potential is given by
\begin{equation}
V(\phi) =\frac{\lambda}{6} \left(\left|\mathcal{H}\right|^2 - \frac{v^2}{2} \right)^2 
=   \frac{\lambda}{24}  \phi^4 + \frac{\lambda}{6}  \phi^3 v + \frac{\lambda}{6}  \phi^2 v^2  \,\, ,
\label{eq-Vtree}
\end{equation}
where we adopt Ford, Jack and Jones conventions for the Higgs quartic coupling $\lambda$ in Ref.~\cite{Ford:1992}.
The tree-level relations for the masses of the Higgs boson and fermions $f$ are given by 
$m_H^2= \lambda v^2 / 3$ and $m_f =   h_f  v/ \sqrt{2}$, where $h_f$ denotes the fermion Yukawa coupling.

To go beyond the tree-level, we adopt the $\overline{\rm MS}$ scheme and extrapolate the Higgs effective potential, $V_{\text{eff}}(\phi)$, 
at very high energies.
We will consider the matching and the renormalization group equations (RGE) evolution of the relevant couplings which, in addition to $\lambda$, are: 
the three gauge couplings, $g$, $g'$, $g_s$, the top Yukawa coupling, $h_t$, and the anomalous dimension of the Higgs field, $\gamma$. 
In the effective potential at large values of $\phi$, potentially large logarithms appear of the type $\log(\phi/\mu)$, where $\mu$ is the renormalization scale.
To not spoil the applicability of perturbation theory, they are treated by means of the renormalization group equations.
For fixed values of the bare parameters, the effective potential must be independent of the renormalization scale $\mu$ \cite{Coleman}.
The formal solution of the RGE is
\beq
V_{\text{eff}}(\mu,\lambda_i,\phi)= V_{\text{eff}}(\mu(t),\lambda_i(t),\phi(t))\,,
\label{eq-Veff}
\eeq
where $i$ stands for each one of the SM couplings $\lambda_i$, $\gamma$ is the background field anomalous dimension and
\beq
\mu(t)=e^t \mu  \, ,\,\, \phi(t) =e^{\Gamma(t)} \phi \,,\,\, \Gamma(t)=- \int_0^t \gamma(\lambda(t')) dt' \, .
\label{eq-mu}
\eeq
The running of the SM couplings, $\lambda_i(t)$, is thus specified by the $\beta$-function equations
\beq
\frac{d \lambda_i(t)}{dt} = \beta_i (\lambda_i(t)) \, ,
\eeq
and subject to the boundary conditions $\lambda_i(0)=\lambda_i$.
We now turn to the issue of the boundary conditions to be given at the low energy matching scale $\mu(0)=\mu$.

\subsection{Matching}
\label{sec-match}

In order to obtain the values of the relevant parameters ($g$, $g'$, $g_s$, $h_t$ and $\lambda$) at the matching scale $\mu(0)$,
we exploit the results of a recent detailed analysis performed by Alam and Martin in Ref.~\cite{Alam:2022cdv},
where $\mu(0)=200$ GeV and the data from the PDG 2022 review are considered \cite{PDG2022}.
While uncertainties in the matching of $g(0)$ and $g'(0)$ are negligible,
the most significant uncertainties are those associated with $g_s(0)$, $h_t(0)$ and $\lambda(0)$. 

Simplified expressions which capture the dominant dependences and sources of uncertainty for the latter three couplings can be found in\,\cite{Alam:2022cdv}, and  
each of them is at present dominated by the experimental error in the related observable.
In this way, the uncertainty in the value of $g_s(0)$ is dominated by the experimental error on the value of the strong coupling constant at $m_Z$ in the SM with five flavors, 
$\alpha_s^{(5)}$; the uncertainties in $\lambda(0)$ and $h_t(0)$ are dominated  by the experimental error respectively on the Higgs mass, $m_H$, 
and the top pole mass, $m_t$. One gets from\,\cite{Alam:2022cdv}:
\beq
g_s (0) \simeq 1.15251 \,\left(1 + 0.00378 \,\frac{\alpha_s^{(5)}- 0.1179} {10^{-3}} \right)\,\,\, ,   
\eeq 
\beq
\lambda(0)/6 \simeq 0.123533 \,\left( 1+  0.001682 \, \frac{m_H- 125.25\,  {\rm GeV}}{ 0.1\, {\rm GeV}}  \right) \,\, \, ,
\eeq
\beq
h_t(0) \simeq 0.923777 \left( 1+ 0.006352  \,\frac{m_t- 172.5\, {\rm GeV}}{1 \,{\rm GeV}}  \right) \,\,\, .
\eeq

In the present work, we use the PDG 2022~\cite{PDG2022} world averages for the mean values and $1\sigma$ errors:
${\alpha}_s^{\rm (5)\,exp}=0.1179$ and $\Delta \alpha_s^{(5)}=0.0009$; $m_H^{\rm exp}=125.25$ GeV and $\Delta m_H =0.17$ GeV;
$m_t^{\rm exp}=172.5$ and $\Delta m_t= 0.7$ GeV. 
Our lower $2\sigma$ and $3 \sigma$ values for $m_t$ are thus respectively $171.1$ GeV and $170.4$ GeV. 

As it is well known, the experimental determination of the top pole mass is a delicate and debated issue.
Considering direct measurements, the PDG 2022 world average in the Top Quark Listings \cite{PDG2022} is $m_t = 172.76 \pm 0.30$ GeV.
In addition to the latter experimental error one should also include the theoretical error taking into account any difference between the top pole mass, 
$m_t$, and the mass parameter implemented in the Monte Carlo event generators employed by the experimental groups, estimated to be
 $\Delta m_t^{MC} = \pm 0.52$ GeV \cite{PDG2022}. 
 The September 2023 ATLAS and CMS combined analysis from data collected at $\sqrt{s}=7$ and $8$ TeV gave $m_t= 172.52 \pm 0.33$ GeV \cite{CMS:2023wnd}.
Considering instead cross-section measurements, the PDG 2022 world average in the Top Quark Listings \cite{PDG2022} is $m_t = 172.5 \pm 0.7$ GeV.
According to the recent analysis by~\cite{Garzelli:2023rvx}, the updated result from cross-section measurements gives a smaller value, $m_t= 171.54^{+ 0.28}_{-0.31}$ GeV; the central value of the latter being about $1.5 \sigma$ below the corresponding PDG 2022 central value.

\subsection{Running}

The $\beta$-functions can be organized as a sum of contributions with increasing number of loops:
\beq
 \frac{d}{d t} \lambda_i(t)=\kappa \beta_{\lambda_i}^{(1)}+\kappa^2  \beta_{\lambda_i}^{(2)} +\kappa^3  \beta_{\lambda_i}^{(3)}  + ...\, ,  \label{eq-RGE}
\eeq
where $\kappa = 1/(16 \pi^2)$ and the apex on the $\beta$-functions represents the loop order.
Here, we are interested in the RGE dependence of the couplings $g$, $g'$, $g_s$, $h_t$, $\lambda$, $\gamma$. 

The one-loop and two-loop expressions for the $\beta$-functions in the SM are well known and can be found, for instance, in Ref.~\cite{Ford:1992}. 
The complete three-loop $\beta$-functions for the SM have been computed in Refs.~\cite{Mihaila:2012,Mihaila1, ChetyrkinZoller,Chetyrkin:2013, BednyakovPikelnerVelizhanin,BednyakovPikelnerVelizhanin1,Bednyakov:2013, Bednyakov:2014}.
The dominant four-loop contribution to the running of the strong gauge coupling has also been computed in Refs.~\cite{Zoller:2015tha,Bednyakov:2015}. 
There are recent results about the four-loop $\beta$-functions:
the leading four-loop QCD contribution to the $\beta$-function of the Higgs self coupling has been derived in \cite{Chetyrkin:2016ruf},
while the full four-loop $\beta$-functions for the gauge couplings have been given in \cite{Dav:2020}.
We checked that, for the sake of our analysis, it is enough to work up to three-loop, as the four-loop contribution has a negligible impact.

\begin{figure}[t!]
 \begin{center}
  \includegraphics[width=7.5cm]{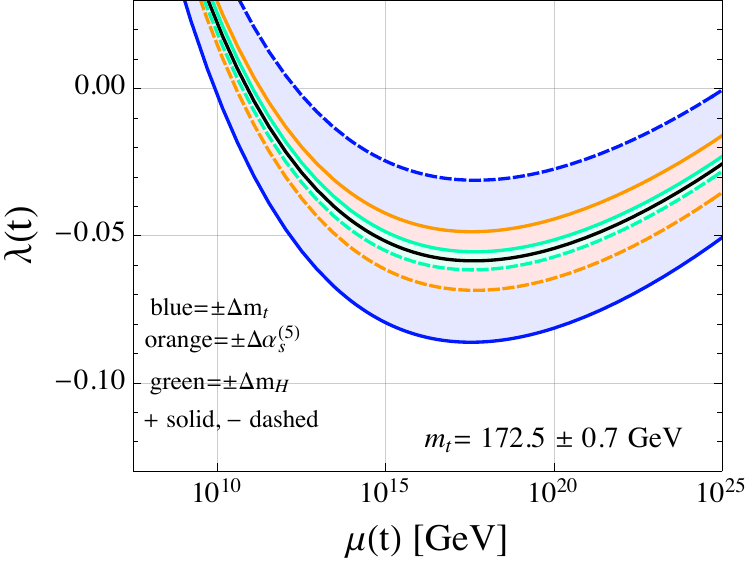}  \,\,\, \, \,\,\, \includegraphics[width=7.5cm]{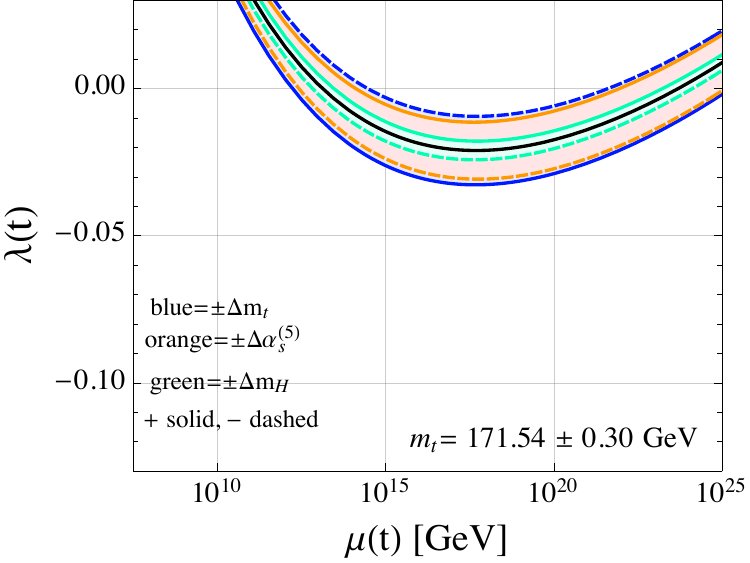}     \end{center}
\caption{\baselineskip=15 pt \small \it
Evolution of $\lambda$ and its experimental variations. Left: We use the PGD 2022 \cite{PDG2022} world averages for the mean values and the $1\sigma$ errors. Right: The same, but adopting the range for $m_t$ suggested in~\cite{Garzelli:2023rvx}.} 
\label{fig-lambda-all}
\vskip .5 cm
\end{figure}

As it is well known, the central values of $m_t$, $\alpha_s^{(5)}$ and $m_H$ suggest that $\lambda$ has a small negative minimum at high energies. 
The left panel of Fig.~\ref{fig-lambda-all} shows the behavior of $\lambda$ for the central values previously mentioned (solid central line),
and its variation associated to the $1\sigma$ range for $m_t$ (blue), $\alpha_s^{(5)}$ (orange) and $m_H$ (green). 
Notice that the variation associated to the $1\sigma$ error in $m_H$ is the less significant one. 
An increase by $+(-) 1\sigma$ in a parameter corresponds to the corresponding solid (dashed) curve. 
Although the central values of the input parameters favor negative values for $\lambda$ above $\mu (t) \sim 10^{11}$ GeV, 
one can see that stability is allowed within $2\sigma$. 
Stability is even more favored adopting the results for $m_t$ according to the analysis by Garzelli et al.~in Ref.~\cite{Garzelli:2023rvx}, as shown in the right plot of Fig.~\ref{fig-lambda-all}.
In the future it might be difficult to exclude stability in a robust way, 
given the challenges associated to a precise determination of $m_t$~\cite{Franceschini:2022veh}.

It is common to identify the instability scale with the scale where $\lambda(t)$ goes negative.
However, the instability scale is, strictly speaking, the renormalization scale where the effective potential turns negative, so that a detailed calculation of the effective potential is relevant.

\subsection{The Higgs effective potential}
\label{sec-Heff}

The RGE-improved effective potential at high field values can be expanded as a sum of tree-level plus increasing loop contributions. 
The one-loop contribution has been computed by Coleman and Weinberg in Ref.~\cite{Coleman:1973jx}; 
the two-loop contribution was derived by Ford, Jack and Jones in Ref.~\cite{Ford:1992} and cast in a more compact form in Refs.~\cite{Degrassi,Buttazzo:2013uya}. 
Recently, there have been additional refinements extending the calculation of the effective potential to three-loop~\cite{Martin:2017lqn} and to the leading four-loop QCD contribution~\cite{Martin:2015eia}.

We adopt here the procedure outlined in~\cite{Andreassen:2014gha}; however (as already done in Refs.~\cite{Iacobellis:2016eof, Masina:2018ejw}, to which we refer for more details), we prefer to work with the wave-function renormalized field, $\phi(t)$, instead of the classical one, $\phi$. Explicitly, in the Landau gauge:
\beq
V_{\rm eff} = V^{(0)} +V^{(1)}  +V^{(2)} +...   \,\,\, ,
\eeq
where
\beq
V^{(0)} = \frac{\lambda(t)}{24}  \phi^4(t)\,\,\,,
\eeq
\bea
V^{(1)}&=& \frac{1}{24} \frac{6}{(4 \pi)^2} 
 \left[  
6   \left(\frac{g^2(t)}{4} \right)^2 \left(  \log \frac{ \frac{g^2(t)}{4} \phi^2(t)}{\mu^2(t)} -\frac{5}{6} \right) \right. \nonumber \\
&+& 3 \left(\frac{g^2(t)+{g^{\prime\,2}(t)}}{4} \right)^2 \left(  \log \frac{ \frac{g^2(t)+{g^{\prime\, 2}(t)}}{4}  \phi^2(t)}{\mu^2(t)} - \frac{5}{6} \right)    \nonumber \\
&-&  \left. 12 \left( \frac{h_t^2(t)}{2}  \right)^2 \left(  \log \frac{\frac{h_t^2(t)}{2}  \phi^2(t)}{\mu^2(t)} -\frac{3}{2} \right) 
 \right]   \phi^4(t) \,,    
\label{eqV1}
\eea
while $V^{(2)}$ can be explicitly found in~\cite{Degrassi,Buttazzo:2013uya}. For large values of the Higgs field $\phi(t)\gg v$ we are only keeping the quartic coupling in the tree-level potential.

The RGE-improved effective potential is gauge dependent.
However the Nielsen identity~\cite{Nielsen:1975fs} guarantees that the effective potential is gauge independent where it is stationary.
This happens for instance for the critical configuration with two degenerate vacua~\cite{Froggatt:1995rt}, and for an inflection point configuration.
On the other hand, the values of the low energy input parameters, as $m_t$, $m_H$ and $\alpha_s^{(5)}$, ensuring stationary configurations are gauge independent~\cite{Iacobellis:2016eof}.
Thus, working in the Landau gauge is perfectly consistent in order to calculate the value of the effective potential at a stationary point, call it $V_s$, or the value of the input parameters providing it. 

Nevertheless, one has to be aware that the truncation of the effective potential loop expansion, at some loop order, introduces an unavoidable theoretical error both in $V_s$ and in the input parameters. 
For this sake, it is useful to introduce the parameter $\alpha$ via $\mu(t)= \alpha \, \phi(t)$ and study how the observables depend on it \cite{Casas:1996aq}.
The higher the order of the considered loop expansion, the milder the $V_s$ dependence on $\alpha$,
as shown explicitly in~\cite{Iacobellis:2016eof} for the cases of two degenerate vacua and a rising inflection point; 
in particular, the value of $\alpha$ for which the one-loop and two-loop (which is almost $\alpha$-independent) expressions for $V_s$ (and the input parameters) agree is $\alpha \simeq 0.3$.
In the following we work at two-loop order taking $\alpha=1$. 

\begin{figure}[htb!]
\vskip .5cm 
 \begin{center}
\includegraphics[width=15cm]{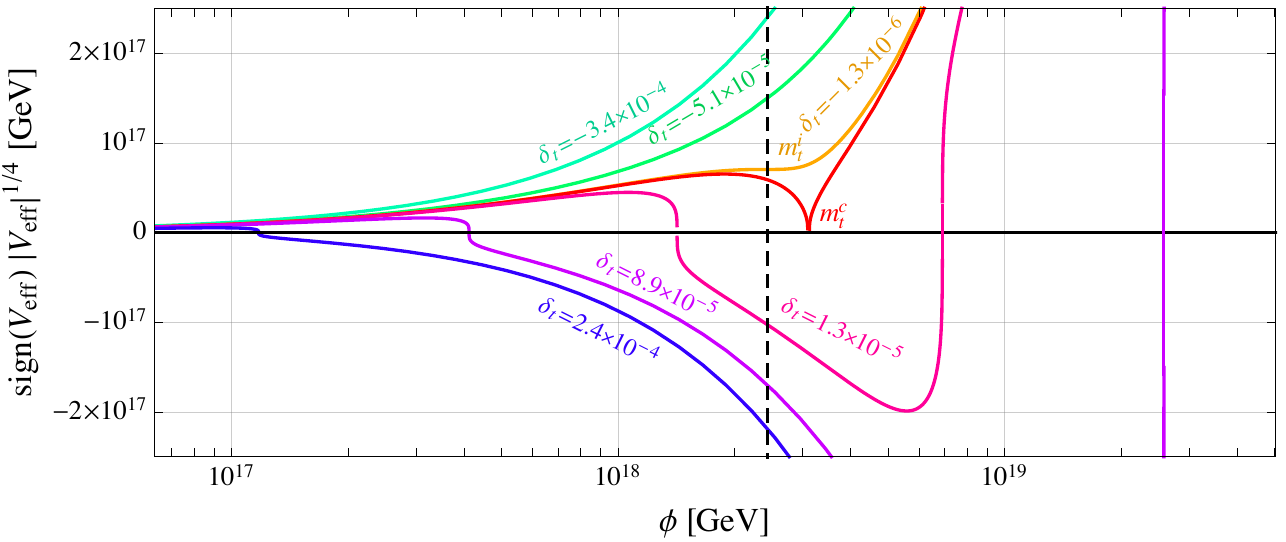}  
 \end{center}
\caption{\baselineskip=15 pt \small \it
Illustrative shapes for the effective potential at high energies. Central values of $m_H$ and $\alpha_s^{(5)}$ are assumed; various values of the top mass are considered, which increase going from top to bottom. The vertical dashed line shows the value of the reduced Planck mass.} 
\label{fig-shapes}
\vskip .5 cm
\end{figure}
Possible shapes of the effective potential are illustrated in Fig.~\ref{fig-shapes},
for central values of $m_H$ and $\alpha_s^{(5)}$ and increasing values of the top mass, going from top to bottom. 
The selected values of the top mass include the critical value $m_t^c$ corresponding to two degenerate vacua and the value corresponding to an inflection point, $m_t^i$. 
Since a tiny variation in the top mass corresponds to a drastic change in the shape of the potential for values of $m_t$ close to $m_t^c$, it is useful to introduce the fractional deviation from criticality, $\delta_t$, as
\beq
m_t = m_t^c \,(1 + \delta_t)  \,\,\, .
\label{deltat}
\eeq

Focussing on the critical configuration (for the central values of $m_H$ and $\alpha_s^{(5)}$ discussed previously and considered in Fig.~\ref{fig-shapes}),
it turns out that: 
$m_t^c \simeq 171.0588$ GeV, close to the lower $2\sigma$ value of the PDG 2022, that is $171.1$ GeV; 
the second degenerate minimum takes place at $\phi_c \approx 3 \times  10^{18}$ GeV, slightly beyond the reduced Planck mass (vertical dashed line);
the maximum of the potential barrier between the electroweak and second degenerate minimum is $V^{1/4}_B \approx 6 \times 10^{16}$ GeV
and takes place at $\phi_B \approx 2 \times 10^{18}$ GeV.
For an inflection point configuration one has $\delta_t \simeq - 1.3 \times 10^{-6}$. 
Notice that the PDG 2022 central and lower $2\sigma$ values of $m_t$ correspond 
respectively to $\delta_t = 8.4 \times 10^{-3}$ and $\delta_t = 2.4 \times 10^{-4}$;
the latter corresponds to the bottom curve in Fig.~\ref{fig-shapes}.

\section{The barrier between the two minima}
\label{sec:barrier}

For $m_t \leq m_t^i$, the Higgs effective potential is stable and monotonously increasing with $\phi$. 
For $m_t \geq m_t^i$, the potential displays a second minimum at high energies. 
We now study in detail the features of the barrier separating the two minima.
This applies to stable ($m_t^i \leq m_t \leq m_t^c $) as well as to metastable ($m_t > m^c_t$) configurations, 
for which the high energy minimum is higher or deeper than the electroweak one, respectively.

\subsection{The barrier for metastable configurations}

\begin{figure}[h]
\vskip .5cm 
 \begin{center} 
  \includegraphics[width=7.95cm]{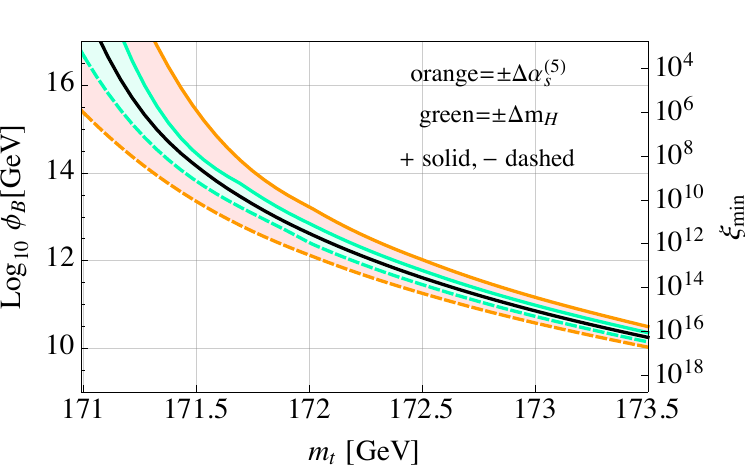}     \,\,\,\,\,
\includegraphics[width=7.45cm]{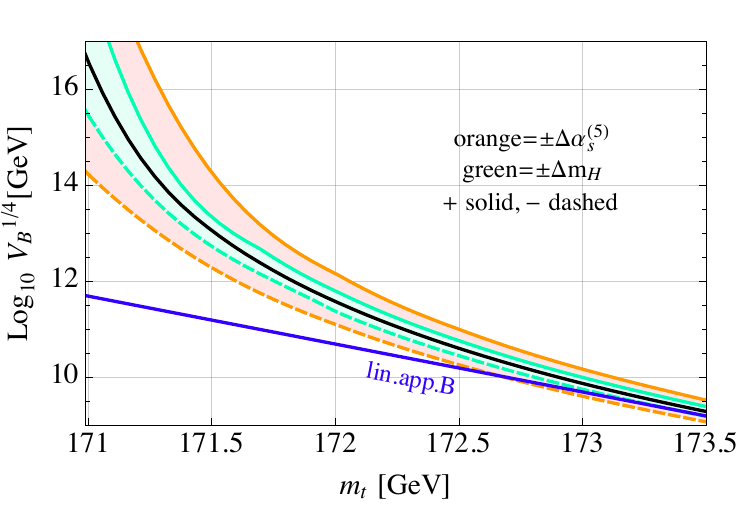}    
 \end{center}
\caption{\baselineskip=15 pt \small \it
The position (left) and the value (right) of the barrier between the electroweak and high energy minima. 
The solid line corresponds to the central values of the inputs according to PDG 2022~\cite{PDG2022}; 
the inner (green) and the outer (orange) bands show the effect of the $1\sigma$ variation in $m_H$ and $\alpha_s^{(5)}$ respectively.
Also shown is the linear approximation of Buttazzo et al.~\cite{Buttazzo:2013uya}.  The quantity $\xi_{min}$ (right vertical axis of the left plot) will be introduced in the following. 
} 
\label{fig-Barr}
\vskip .5 cm
\end{figure}

The (gauge non-invariant) field value corresponding to the top of the barrier, $\phi_B$, is shown in the left plot of Fig.~\ref{fig-Barr} as a function of $m_t$. 
The central line corresponds to the mean values of the inputs as discussed previously, and the band shows the impact of the $1\sigma$ experimental errors.
As for the instability scale, one defines the parameter $t_V$ by means of the constraint $V_{eff}(t_V)=0$. 
The corresponding value of the Higgs field is $\phi(t_V)$ (which is not gauge invariant); it turns out that $\phi(t_V) =1.29 \, \phi_B$.

The value of the Higgs potential at the top of a barrier, $V_B$, is a gauge invariant quantity, as it corresponds to a stationary configuration.
We plot it in the right panel of Fig.~\ref{fig-Barr}, as a function of $m_t$. 
To the best of our knowledge, the dependence of $V_B$ on $m_t$ has not been studied in such detail so far.
Previously, a linear approximation was derived by Buttazzo et al.~\cite{Buttazzo:2013uya} (see also~\cite{Franceschini:2022veh}),
which is included for comparison in the plot; such approximation was derived in view of large values of $m_t$, 
so it fails for $m_t \lesssim 172.5$ GeV. 

The magnitude of $V_B$ is relevant in view of applications of the Higgs potential in the context of Higgs inflation. 
It is tempting to precisely exploit the Higgs potential for inflation, and interpret the Higgs directly as the inflaton. 
There are already constraints on this hypothesis: from the CMB data one can derive an upper bound on the potential responsible for inflation. 
We briefly review this argument later in this section.

\subsection{The barrier for stable configurations}
\label{sec-two}

We now focus on stable configurations, \textit{i.e.}~from the configuration with two degenerate vacua to the configuration with an inflection point.
Once $m_H$ and $\alpha_s^{(5)}$ have been fixed, the critical value of $m_t$ for which the SM displays two degenerate vacua, $m^c_t$, 
is a gauge invariant quantity. 
This value has been studied in many works, often showing the critical top mass line (separating the stability and metastability regions) as a function of the Higgs mass~\cite{Degrassi, Masina:2012tz, Buttazzo:2013uya, Bednyakov:2015sca}.

\begin{figure}[h]
\vskip 0.5cm 
 \begin{center}
\includegraphics[width=7.6cm]{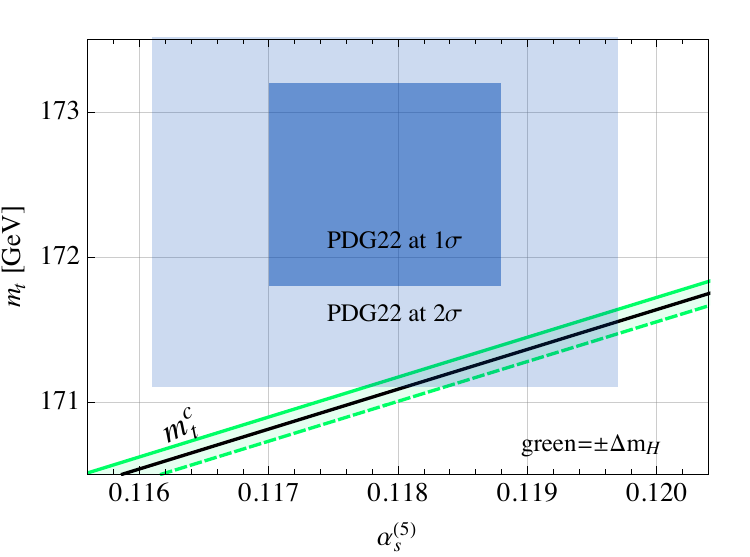} \\ \vskip .4cm 
\includegraphics[width=7.6cm]{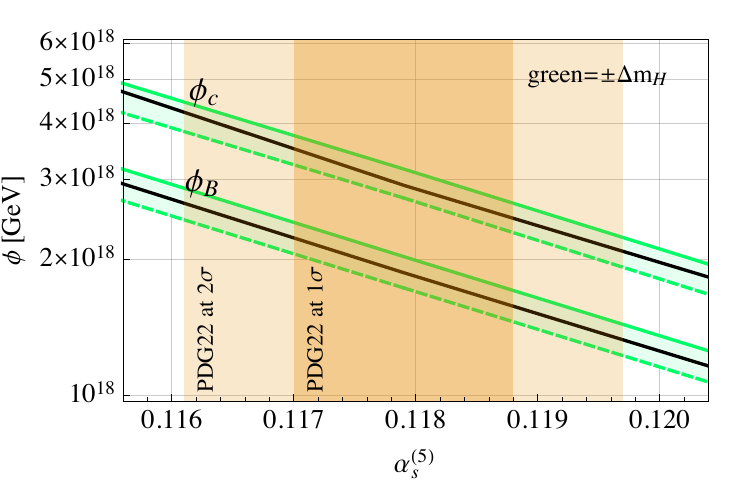}  \,\,\, \,
\includegraphics[width=7.6cm]{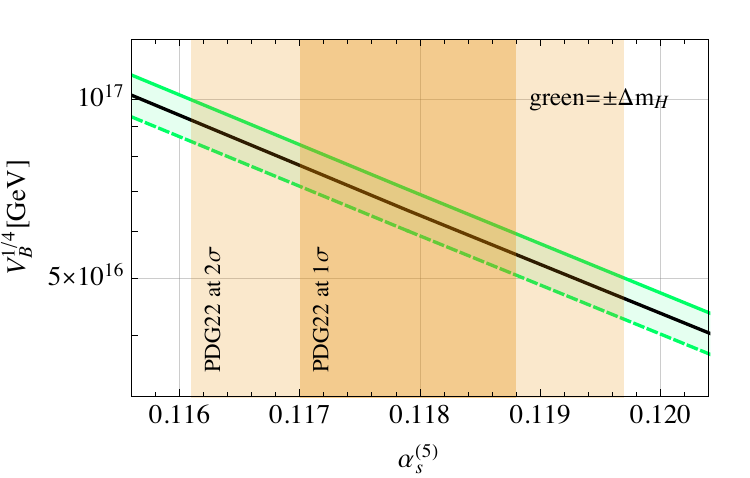}
\end{center}
\caption{\baselineskip=15 pt \small \it Two degenerate vacua (critical configuration). 
As a function of $\alpha_s^{(5)}$, we display $m_t^c$ (top panel), $\phi_c$ and $\phi_B$ (bottom left panel), $V_B^{1/4}$ (bottom right panel). 
The solid lines correspond to the central values of the inputs according to PDG 2022~\cite{PDG2022}; the (green) band shows the impact of the $1\sigma$ experimental error in $m_H$. The shaded (blue and orange) regions display the $1\sigma$ and $2\sigma$ ranges for $m_t$ and $\alpha_s^{(5)}$ respectively.}   
\label{fig-c}
\vskip .5 cm
\end{figure}

Given the impressive precision achieved on $m_H$, it is by now more useful to plot the critical top mass line as a function of $\alpha_s^{(5)}$, 
as done in the upper panel of Fig.\,\ref{fig-c}, where we can see the mild variation induced by varying $m_H$ within $1\sigma$. 
The shaded rectangles display the $1\sigma$ and $2\sigma$ variations in $m_t$ and in $\alpha^{(5)}_s$, according to the PDG 2022.
The criticality line is inside the lower right corner of the $2\sigma$ rectangle \footnote{Our results agree with the recent work~\cite{Hiller:2024zjp}.}.
This shows that stability is at present compatible with experimental data, and that it will be challenging to robustly support metastability in future~\cite{Franceschini:2022veh}.

In the left bottom panel of Fig.\,\ref{fig-c} we display, always for a critical configuration, the value of $\phi_c$, where the high energy minimum takes place
(notice that it can be slightly beyond the reduced Planck scale, which is about $2.4\times 10^{18}$ GeV). 
The barrier separating the electroweak and the second degenerate minimum at $\phi_c$ has a maximum at $\phi_B$, which is also
displayed in the same panel (and turns out to be slightly lower than the reduced Planck scale).
The height of the barrier, $V_B^{1/4} $, is displayed in the right bottom panel of Fig.\,\ref{fig-c}; 
it turns out to be just below the Planck scale.
Notice that $V_B^{1/4}$, $\phi_c$ and $\phi_B$ all decrease when increasing $\alpha_s^{(5)}$.
As can be seen from the plots, the central values for the parameters characterizing a critical configuration are: 
$\phi_c \approx 3 \times 10^{18}$ GeV, $\phi_B \approx 2 \times 10^{18}$ GeV, 
$V^{1/4}_B \approx 6 \times 10^{16}$ GeV. 
The latter scale cannot be much reduced by changing the input parameters: considering the upper $2\sigma$ value of $\alpha_s^{(5)}$, 
one cannot achieve a value smaller than $V^{1/4}_B \approx 4.5 \times 10^{16}$ GeV.

Notice that, for any value of the top mass in the range $m_t^i \leq m_t \leq m_t^c $ (from a degenerate second minimum to the case of an inflection point configuration), 
the values of $\phi_B$ and $V_B$ are practically the same as those indicated in Fig.\,\ref{fig-c}.

\subsection{Constraints on the height of the barrier from CMB}

In the following, we 
review the constraints provided by the CMB on the inflationary potential, $U_i$, in view of a possible identification of the Higgs with the inflaton field. 
We introduce a bar over a quantity to indicate that it is written in units of the reduced Planck mass,
$M_{P}\simeq 2.43\times10^{18} \,{\rm GeV}$,
and introduce the subscript $i$ for a quantity to be calculated at the beginning of inflation, say $50-60$ e-folds before its end.
According to the PDG 2022 review on cosmological parameters~\cite{PDG2022}, 
the observed amplitude of the density fluctuations is given by
\beq
2.1 \times  10^{-9} \approx \Delta_R^2 =\frac{\bar H_i^2}{8 \pi^2 \epsilon_i}  = \frac{2 \bar U_{i}}{3 \pi^2 r}  \,\,\,  ,
\eeq
where in the last equality we exploited the relation $\bar H_i^2 =\bar U_{i}/3$, where $H_i$ is the Hubble constant, and assumed a slow-roll phase 
(with slow-roll parameter $\epsilon_i$) for which the tensor-to-scalar ratio is $r =16\,\epsilon_i$. The constraint on the inflationary potential is thus
\beq
\bar U_{i} ^{1/4}  \equiv \frac{ U_{i} ^{1/4}}{M_P}=\left(  \frac{3}{2} \pi^2 \,r \, \Delta_R^2 \right)^{1/4} \approx  
1.3 \times 10^{-2} \,\, r^{1/4}  \,\, .
\label{eq-UiCMB}
\eeq
The most recent analysis of the BICEP, Keck Array and Planck data~\cite{BICEP:2021xfz} places an upper bound on the tensor-to-scalar ratio $r < 0.036$ at the 95\% CL, so that from the equation above one gets $U_{i}^{1/4} <  2.5 \times 10^{16}$ GeV.
Assuming that $r$ is not far from being detected, the scale of $U_i^{1/4}$ should correspond to the GUT scale. 

It is worth emphasizing that the previous quantities like $\phi_c$, $\phi_B$ and $V_B$ do depend just on the SM parameters, and have nothing to do with gravity.
For metastable configurations~Fig.\,\ref{fig-Barr} shows for which values of the input parameters $V_B^{1/4}$ is in the GUT range:
this requires low values of $m_t$, close to the $2 \sigma$ lower range.
For $m_t \lesssim 171.5$ GeV, the values of $V_B^{1/4}$ can even exceed the GUT scale. In order to be exploited for inflation,  
the Higgs potential would thus require a modification, as for instance a flattening by means of a non-minimal coupling to gravity. 

As it is well known (and will be reviewed in the following), in the metric formulation the Higgs inflation model predicts $r=12/N^2=0.0033-0.0048$ for $N=60-50$, so that CMB data would suggest the value $U_{i}^{1/4} =  (7.8-8.5) \times 10^{15}$ GeV
(or equivalently, $\bar U_{i} ^{1/4} \approx 10^{-2.5}$ as shown in Fig.~\ref{fig-P1}). 
For values of $m_t \gtrsim 171.5$ GeV, the values of $V_B^{1/4}$ are smaller than the GUT scale.
The introduction of a non-minimal coupling to achieve Higgs inflation is thus useless in this case, as the flattened potential would stay in any case below the 
desired value, $U_{i}^{1/4} =  (7.8-8.5) \times 10^{15}$ GeV. Other modifications of the SM Higgs potential would be required in view of inflation, as for instance: 
\textit{i)} introducing new physics to increase $V_B^{1/4}$, like adding another scalar field \cite{EliasMiro:2012ay,Cado:2022evn} or in $R^2$-Higgs inflation~\cite{Cado:2023zbm}; 
\textit{ii)} considering a model with very small $r$, allowing for values of $V_B^{1/4}$ much smaller than the GUT scale (see~\cite{Yin:2022fgo} for a related attempt).
For a stable configuration instead, as stressed previously while commenting Fig.~\ref{fig-c}, one has $V_B^{1/4} \gtrsim 4.5\times 10^{16}$ GeV,
which is in tension with the constraint from CMB, $U_{i}^{1/4} <  2.5 \times 10^{16}$ GeV.
Models with a shallow vacuum \cite{MasinaHiggsmass,Masinatop,Masinahybrid,Masina:2014yga} and inflection point Higgs inflation (also called critical Higgs inflation) \cite{Hamada:2014iga, Masina:2014yga, Bezcritical, Hamada:2014wna, Ezquiaga:2017fvi, Salvio:2017oyf} are thus in tension with CMB data \cite{Masina:2018ejw}, 
unless invoking modifications to the Higgs effective potential \cite{Salvio:2018rv}.

Finally, in the Palatini formulation the value of $r$ depends on $\xi$ and is much smaller than in the metric formulation, as we will see, leading to much lower values of $\bar U_i$. This will allow going to larger values of $m_t$ consistently with cosmological observables.

\vskip 1.cm
\section{The Higgs potential with a non-minimal coupling}
\label{sec:potential}

We now modify the Higgs potential by introducing a non-minimal gravitational coupling $\xi$ between the SM Higgs doublet $\mathcal{H}$ and the Ricci scalar $R$~\cite{BezHiggs}. 
The classical action for Higgs inflation is
\begin{equation}
\mathcal{S}=\int d^{4}x \, \sqrt{-g}\left[\mathcal{L}_{SM}-\frac{M^{2}}{2} R -\xi\left|\mathcal{H}\right|^{2} R\right]\,,
\label{eq-act}
\end{equation}
where $g$ is the determinant of the FLRW 
metric, 
$\mathcal{L}_{SM}$ is the SM Lagrangian and $M$ is a parameter with the dimension of a mass,
to be identified with the reduced Planck mass, $M_{P}=1/(8\pi G_{N})^{1/2}$, 
where $G_N$ is the Newton constant, 
in order to obtain the equations of general relativity.

In this section we will consider both the metric and the Palatini formalisms of gravity~\cite{Palatini:1919ffw}. In the metric formulation the connection is identified with the Levi-Civita connection which is written in terms of the metric, while in the Palatini formalism the connection and metric are independent quantities in the Jordan frame. In the Einstein frame the connection is always equal to the Levi-Civita connection but there is a difference in the kinetic term of the field which is coupled to the Ricci scalar. We will thus consider here a general formulation where the difference between both formalisms is characterized by a parameter $\kappa$.
 
From a cosmological point of view, the relevant part of the action above is
\begin{equation}
\mathcal{S}_{J}=\int d^{4}x \, \sqrt{-g}\left[  (\partial_{\mu}\mathcal{H})^{\dagger}(\partial^{\mu}\mathcal{H})-\frac{M_{P}^{2}}{2} R-\xi\left|\mathcal{H}\right|^{2} R-V\right]\,,
\end{equation}
where $V$ is the SM potential of Eq.~(\ref{eq-Vtree}) and the subscript $J$ means that the action is evaluated in the Jordan frame.
In order to remove the non-minimal coupling we introduce a conformal (or Weyl) transformation:
\begin{equation}\label{weyl}
\tilde{g}_{\mu\nu}=\Omega^{2}g_{\mu\nu},\qquad \Omega^{2}\equiv1+2\xi\frac{\left|\mathcal{H}\right|^{2}}{M_{P}^{2}} = 1+ \xi \bar \phi^2\,.
\end{equation}
where the last equality holds in the Higgs representation $\mathcal{H}=e^{i\vec\chi\cdot\vec\sigma}(0,(\phi+v)/\sqrt{2})^T$, where $\vec\chi$ are the Goldstone bosons and $\phi$ is the radial mode. 
The Einstein frame action is obtained when gravity is canonically normalized,
\begin{equation}\label{acteinsim}
\mathcal{S}_{E}=\int d^{4}x\, \sqrt{-\tilde{g}}\left[K\frac{(\partial\phi)^{2}}{2}-\frac{M_{P}^{2}}{2}\tilde{R}-\frac{V}{\Omega^{4}}\right]\,\,\,,\,\,
K = \frac{ \Omega^{2}+ \frac{3}{2} \kappa \left(   \frac{d\Omega^2}{d \bar \phi} \right)^2  }{\Omega^{4}}  \,\, ,
\end{equation}
where $\kappa=1$ in the metric formulation, while $\kappa=0$ in the Palatini formulation.

The kinetic term for the classical Higgs field $\phi$ is made canonical by the redefinition $\phi=  \phi(\chi)$ such that
\begin{equation}
\frac{d \chi}{d \phi}= \sqrt{K}=
\frac{\sqrt{ 1+\xi  \bar \phi^2 + 6 \kappa (\frac{1}{2} \frac{d\xi}{d \bar \phi}  \bar \phi+ \xi)^2   \bar  \phi^2} } { 1+ \xi \bar  \phi^2}\,,\qquad \left.\chi( \phi)\right|_{\phi=0}=0\,\, ,
\label{eq-chiphi}
\end{equation}
so that the final expression for the Einstein frame action and the potential $U$ felt by $\chi$ are
\begin{equation}
\mathcal{S}_{E}=\int d^{4}x \, \sqrt{-\tilde{g}}\left[ \frac{(\partial\chi)^{2}}{2}-\frac{M_{P}^{2}}{2}\tilde{R}-U(\chi)\right]\,\,\,   , \,\, \,  
U (\chi)= \frac{V}{\Omega^{4}} \simeq \frac{\lambda}{24}\frac{\phi^4(\chi)}{(1+\xi\bar{\phi}^2(\chi))^{2}}\,\, \,\,\, .
\end{equation}
The potential is flat for large field values, $ \phi > M_P/\sqrt{\xi} $, and can in principle provide a slow-roll inflationary phase.

\subsection{Radiative corrections}

We turn to consider the inclusion of radiative corrections: the running of the couplings, now also including the running of the non-minimal coupling $\xi$, 
and the loop corrections to the effective potential.  

The expressions for the $\beta$-functions of the relevant SM couplings, including $\xi(t)$, 
can be found \textit{e.g.}~in Refs.~\cite{Allison, George:2015nza}. 
The non-minimal coupling affects the running through the appearance of a factor $s$ that suppresses the contribution of the physical Higgs 
to the RGEs~\cite{DeSimone:2008ei, Allison}:
\begin{equation}
s(\phi(t))=\frac{1+\xi(t) {\bar \phi^2(t)}}{1+(1+6\xi(t)) \xi(t) {\bar \phi^2(t)}}\,.
\end{equation}
For small field values $\bar \phi(t) \ll 1/{\xi(t)}$, $s\simeq1$, recovering the SM case; in the inflationary regime 
$\bar \phi(t)\gg 1/{\xi(t)}$,  the RGEs differ from those of the SM as quantum loops involving the Higgs field are suppressed by $s\simeq 1/(1+6\xi(t))$.

The RGE-improved effective potential is given by  
\beq
{U}_{\text{eff}}={U}^{(0)}+{U}^{(1)}+{U}^{(2)}+... \,,
\eeq
 with all the involved running couplings evaluated at some renormalization scale $\mu(t)$, conveniently chosen in order to minimize the effect of the logarithms, for the benefit of perturbation theory. 
 
 We will then consider the quantum corrections to the effective potential~\cite{Allison, Fumagalli:2016lls} after the transformation\,(\ref{weyl}) has been done, \textit{i.e.}~in the Einstein frame~\cite{BezHiggs}, where gravitons have canonical kinetic term~\footnote{We will neglect quantum gravity effects, although they might be important at the scales where inflation will take place. 
 }. 
 First, the tree-level RGE-improved potential is rewritten in the Einstein frame, giving at leading order
 \begin{equation}
{U}^{(0)}= \frac{\lambda(t)}{24}  \frac{ { \phi^4(t)}} {\Omega^4(t)} \,\,\,, \,\,\Omega^2(t)=1+\xi(t) \, {\bar \phi^2(t)}\,\,.
\label{treeP}
\end{equation}

As for the one-loop potential, the relevant terms for the gauge bosons $V$ and fermions $\psi$ in the tree level action are given in the Jordan frame by
\begin{equation}
S_J=\int d^4x \sqrt{-g}\left[-\frac{1}{4}  \textrm{tr }F_{\mu\nu}F^{\mu\nu} -\frac{1}{2} \frac{M^2(v)}{v^2} \phi^2 V_\mu V^\mu 
+\bar\psi i D_\mu \gamma^\mu \psi -\frac{\lambda_\psi}{\sqrt{2}}\bar\psi_L \phi \psi_R +\dots \right]
\label{SJ}
\end{equation}
and in the Einstein frame by
\begin{equation}
S_E=\int d^4x \sqrt{-\tilde g}\left[-\frac{1}{4}  \textrm{tr }F_{\mu\nu}F^{\mu\nu} -\frac{1}{2} \frac{M^2(v)}{v^2} \frac{\phi^2}{\Omega^2} V_\mu V^\mu 
+\bar\chi i D_\mu\tilde \gamma^\mu \chi -\frac{\lambda_\psi}{\sqrt{2}}\bar\chi_L \frac{\phi}{\Omega} \chi_R +\dots\right]
\label{SE}
\end{equation}
where $\chi\equiv \Omega^{-3/2}\psi$ and $\tilde\gamma^\mu\equiv \Omega^{-1} \gamma^\mu$ satisfies the anticommutation relation $\{\tilde\gamma^\mu,\tilde\gamma^\nu\}=2\tilde g^{\mu\nu}$. In these  expressions we are neglecting the variation of $\Omega(\phi)$ which is motivated during inflation when slow-roll conditions are fulfilled for the inflaton $\phi$. Notice that the kinetic term of gauge bosons is scale invariant, so gauge bosons do not need any field redefinition.

The one-loop corrections take the form of Eq.~(\ref{eqV1}), with field dependent particle masses computed from
the tree-level action above in the Einstein frame~(\ref{SE}); 
for instance, the contribution of the $W$ to the effective potential is a function of the $W$ mass, $m_W= g \phi/2$.
The two-loop radiative corrections ${U}^{(2)}$ can be found in the same way, operating on the explicit form given in 
Refs.~\cite{Degrassi,Buttazzo:2013uya}.
The appropriate scale for minimizing the effect of the logarithms is given by $\phi(t)/\Omega(t)$. Following the argument illustrated in the previous section, we define the parameter $\alpha$ via $\mu(t)=\alpha \phi(t)/\Omega(t)$~\footnote{It was proposed in Ref.~\cite{Barvinsky} to compute quantum effects in the Jordan frame, so that radiative corrections are evaluated directly from masses provided by the Jordan frame, Eq.~(\ref{SJ}), before the conformal transformation: they are thus given by $V^{(1)}$ of Eq.~(\ref{eqV1}). In this case the scale factor $\mu_J(t)$ in the Jordan frame is related to the scale factor in the Einstein frame $\mu(t)$ by the anomaly relation $\mu_J(t)=\mu(t) \Omega(t)$, providing the manifestation of the conformal anomaly in both frames~\cite{Bezrukov:2010jz}. Even if the choice of the renormalization scale is relevant when a truncated series in perturbation theory is considered, as the former is not physical, and the effective action is renormalization scale invariant, the physical observables do not depend on the renormalization scale choice.}, which give for the one-loop potential
\bea
U^{(0)}+U^{(1)}&=& \frac{1}{24}  \left(   \lambda(t)     +  \frac{6}{(4 \pi)^2} 
 \left[  
6   \left(\frac{g^2(t)}{4} \right)^2 \left(  \log \frac{ g^2(t)}{4 \alpha^2}   -\frac{5}{6} \right) \right.  \right. \nonumber \\
&+& 3 \left(\frac{g^2(t)+{g^{\prime\,2}(t)}}{4} \right)^2 \left(  \log  \frac{g^2(t)+{g^{\prime\,2}(t)}}{4 \alpha^2}   - \frac{5}{6} \right)    \nonumber \\
&-&  \left.  \left.12 \left( \frac{h_t^2(t)}{2}  \right)^2 \left(  \log \frac{h^2_t(t)}{2 \alpha^2}  -\frac{3}{2} \right) 
 \right] \right) \frac{  \phi^4(t)}{ (1+\xi(t) \, {\bar \phi^2(t)})^2 } \,.
\label{eqU1c}
\eea

Notice that the explicit dependence on the field values $\phi(t)$ has disappeared as we are making the choice for the scale $\mu(t)$
\beq
 \mu(t)=\alpha\,  \frac{\phi(t)}{\sqrt{1+ \xi(t) \bar \phi^2(t)}} \,.
 \eeq
For small field values, $\bar \phi(t) \ll 1/\sqrt{\xi(t)}$, we get the same result as in this case $\Omega \approx 1$, so that:
\beq
t=\log \left(   \frac{\alpha \phi(t) }{\mu(0)} \right) \,\,.
\eeq
For large field values, $\bar \phi(t) \gg 1/\sqrt{\xi(t)}$, the situation changes as
$t$ approaches a nearly constant value, approximately given by 
\beq 
t_{\rm lim} \approx \log \left( \frac{ \alpha}{\sqrt{\xi(t_{\rm lim})} }\frac{M_{P}}{\mu(0)} \right)    \,\,\,,
\eeq 
and hence so do the couplings $g(t), g'(t), \lambda(t)$, etc.
As a result, the effective potential approaches a constant value in the inflationary region, even after including radiative corrections~\footnote{The fact that the one-loop corrections do not spoil the flatness of the tree-level potential was already noticed in Ref.~\cite{BezHiggs}.}. 
The constant limit for the potential is thus expected to be 
\beq
U_{\rm lim} \approx \frac{1}{24} \lambda(t_{\rm lim}) \frac{M_P^4 }{\xi^2(t_{\rm lim})} \,\,.
\label{eq-Ulim}
\eeq

In the following, we study how to modify the SM potential by exploiting the flattening induced by the non-minimal coupling $\xi$.
As discussed above, the transition from small to large field values of $\bar \phi(t)$ happens at $1/\sqrt{\xi(t)} \approx 1/\sqrt{\xi(0)} \equiv 1/\sqrt{\xi} \equiv\bar \phi_\xi$, as the running of the parameter $\xi(t)$ is very slow and one can safely neglect it.
It will thus be important to compare $\bar \phi_\xi$ with the parameters characterizing the potential, especially with $\bar \phi_B$.
It is also important to understand the size of the theoretical error associated to the truncation of the effective potential at a certain loop order. 
This error can be estimated by varying $\alpha$, as done in \cite{Iacobellis:2016eof}. 
In the following we thus use the one-loop approximation with $\alpha=0.3$, as it gives nearly the same results as the two-loop one, which we have checked to essentially be $\alpha$-independent.
In particular the critical value of the top quark mass is given by $m_t^c=171.0549$ GeV. Notice that this is just $4$ MeV smaller than the value mentioned 
in Sec. \ref{sec-Heff}, which was obtained by working at two-loop and taking $\alpha=1$. The theoretical uncertainty in our determination of the top mass is thus $\Delta_{th} m_t \simeq 4$ MeV. 
To illustrate the different behaviors of the effective potential we will consider the benchmark configurations that are shown in Tabs.~\ref{table1} and \ref{table2}.

\begin{table}[h]
\begin{center} 
\begin{tabular}{|| c || c| c| c|  c|| } 
\hline
Point & $P_{1a}$ &$P_{1b}$ & $P_{i}$ & $P_{c}$   \\ \hline
$\delta_t$  & -$3.8 \cdot 10^{-3}$ &-$1.5 \cdot 10^{-3}$ &-$1.4  \cdot 10^{-6}$ &$0$ \\ \hline
$m_t$ [GeV]  & 170.40 & 170.80 & 171.0547 & 171.0549 \\ \hline
\end{tabular}
\caption{\it Benchmark points used in the stability region. The value of $\delta_t$ parameterizes the deviation with respect to the critical configuration $P_c$, see Eq.~(\ref{deltat}), for which $\delta_t=0$. }
\label{table1}
\end{center}
\end{table}

\begin{table}[h]
\begin{center} 
\begin{tabular}{|| c || c| c| c|  c| c| c| c|c|c|| } 
\hline
Point & $P_{2a}$ & $P_{2b}$ & $P_{3a}$ &$P_{3b}$ &$P_{3c}$ &$P_{3d}$  \\ \hline
$\delta_t$  & $6.2  \cdot 10^{-6}$ &$1.5  \cdot 10^{-4}$ & $2.6  \cdot 10^{-4}$&$8.5  \cdot 10^{-4}$ & $1.4  \cdot 10^{-3}$& $2.8  \cdot 10^{-3}$\\ \hline
$m_t$ [GeV]  &  171.056 & 171.08 & 171.10 & 171.20  & 171.29 & 171.54\\ \hline
\end{tabular}
\caption{\it Benchmark points used in the metastability region. }
\label{table2}
\end{center}
\end{table}

\subsection{The canonical field and its equation of motion}

As a last step, we have to generalize the relation between $\phi$ and the canonical field $\chi$ to the case of running couplings. As discussed in Refs.~\cite{Espinosa:2015qea, Fumagalli:2016lls}, the kinetic term for the wave-function renormalized Higgs field, $\phi(t)$, can be made canonical by defining the field $\chi (\phi(t))$ by means of a generalization of Eq.~(\ref{eq-chiphi}).
We numerically integrate such equation as done in Ref.~\cite{Masina:2018ejw}, to which we refer for more details.
Once the shape of $U(\chi)$ is known, it is possible to wonder about primordial inflation. 
One has to distinguish between the possibilities of single and multi-field inflation.

In the case of single-field inflation, the Higgs field itself plays the role of the inflaton and,
as it is well known, the inflationary observables are calculated as follows.
By identifying $t$ with the cosmological time~\footnote{Not to be confused with the RGE scale $t$.}, 
the equation of motion of the field $\chi(t)$ is given by
\beq
\chi''(t) +3 \,H(t)\, \chi'(t) = - \frac{dU}{d\chi}(\chi(t)) \,\,, \,\, H^2(t)=  \frac{1}{3M_P^2} \left(U(\chi(t)) + \frac{1}{2} \chi^{\prime\,2}(t) \right) \,\, ,
\eeq
where the initial conditions are $\chi(t_0)=\chi_0$, $\chi'(t_0)=\chi'_0$, and $t_0$ is some initial time.
The time duration of the inflationary phase is represented by the number of e-folds,
\beq
N=\int_{t_i}^{t_e} dt \, H(t)\,\,,
\eeq
where $t_e$ is the time at the end of inflation and $t_i>t_0$ is the time when the inflationary CMB observables, like the density fluctuations $\Delta_R$, the spectral index $n_s$ and the tensor-to-scalar ratio $r$, are measured. 
It is known that $t_i$ should be such that $N \approx 50-60$.

In the case of multi-field inflation instead, 
the Higgs field is not the inflaton 
and one introduces another scalar field, acting as inflaton, so that the Hubble parameter is dominated by the potential of the latter field.

\section{Avoiding metastability by a non-minimal coupling}
\label{sec:flattening}

Studies of the Higgs potential in de Sitter background, during the cosmological inflation period triggered by some other inflaton field, shows that the minimal probability for the Higgs to overcome the top of the barrier after $N$ e-folds of inflation is~\cite{Espinosa:2015qea}
\beq
p(\phi>\phi_B)\simeq 1-\textrm{erf}\left(\frac{\sqrt{2}\pi \phi_B}{\sqrt{N}H}  \right) \,\,\,,
\eeq
where erf$(x)$ is the error function. Imposing that this does not happen in any of the $e^{3N}$ causally disconnected regions that are made during inflation, \textit{i.e.}~$p(\phi>\phi_B)<e^{-3N}$ imposes the condition 
\beq
H_e\lesssim 0.04\, \phi_B \,\, ,
\label{HephiB}
\eeq
where $H_e$ is the Hubble parameter at the end of inflation, and the value of $\phi_B$ for different values of $m_t$ is plotted in the left panel of Fig.~\ref{fig-Barr}. This imposes a strong bound on the value of the Hubble parameter during the inflationary period and, in particular, excludes the case of high-scale inflation. For instance for the central experimental value of the top quark mass, for which $\phi_B\simeq 3 \times 10^{11}$ GeV, the scale of inflation is bounded by $H_e\lesssim  10^{10}$ GeV.
\newline\indent
Of course if the Higgs field is identified with the inflaton, the existence of a metastable minimum precludes inflation unless $\phi_i\lesssim \phi_B$, 
$\phi_i$ being the Higgs field at the beginning of inflation.

A non-minimal coupling to gravity can naturally allow to avoid such metastable minimum (for a pioneering approach to this idea, though with different results than ours, see \cite{Yin:2022fgo}).
if $\xi$ is sufficiently large, so that $1/\sqrt{\xi} = \bar \phi_\xi \lesssim \bar \phi_B$. 
The lower bound on the non-minimal coupling is thus
\beq
\xi_{\rm min} \equiv \frac{1}{{\bar \phi_B}^2} \lesssim \xi  \,\,\, ,
\eeq
which we display in the right vertical axis of the left panel of Fig.~\ref{fig-Barr}. 
Clearly, very large values of $\xi$ are required to avoid metastability, unless $m_t$ is close to the PDG 2022 lower $2\sigma$ range, $m_t=171.1$ GeV~\cite{PDG2022}. 

\begin{figure}[t!]
 \begin{center} 
\vskip .5cm 
  \includegraphics[width=7.5cm]{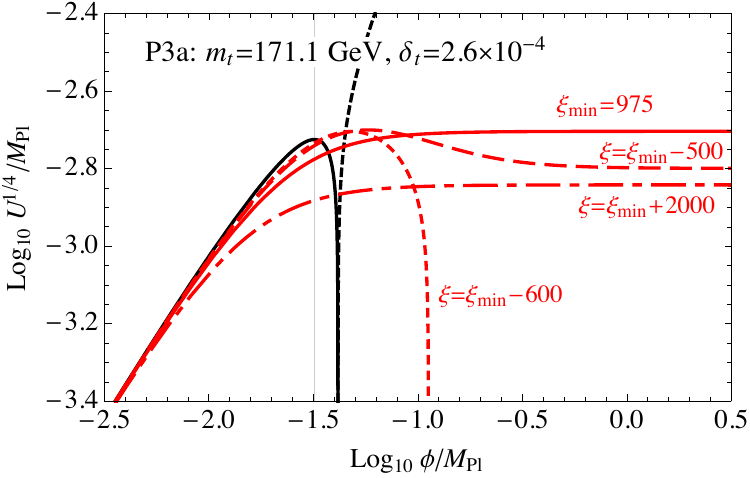}     \,\,\,\,  \includegraphics[width=7.5cm]{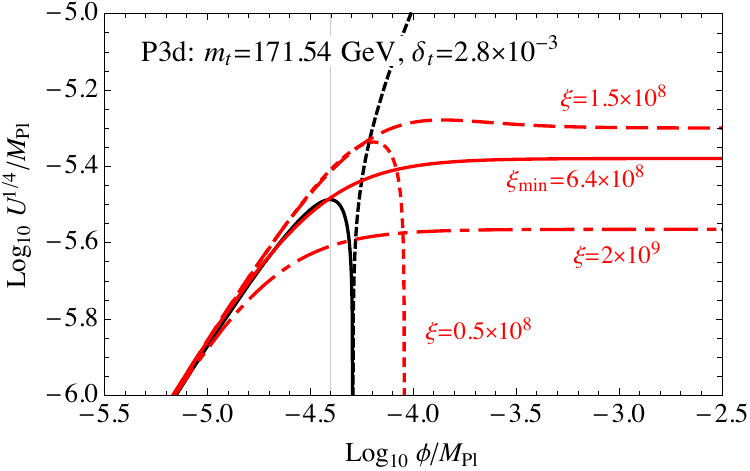} 
   \end{center}
\caption{\baselineskip=15 pt \small \it
Left panel: study of the configurations with $m_t=171.1$ GeV, for the indicated values of $\xi$. The solid line refers to $\xi=\xi_{min}=975$; the short and long dashed lines to $\xi= \xi_{min} -600$ and $\xi= \xi_{min} -500$ respectively; the dot-dashed line to $\xi= \xi_{min} +2000$. 
Right panel: study of the configurations with $m_t=171.54$ GeV, as suggested in \cite{Garzelli:2023rvx}, for the indicated values of $\xi$, including $\xi_{min}=6.4\times 10^8$.}
\label{fig-nonmin}
\vskip .5 cm
\end{figure}

As an illustrative example, we precisely fix $m_t=171.1$ GeV, case $P_{3a}$ in Tab.~\ref{table2}, take central values for $m_H$ and $\alpha_s^{(5)}$ given by the PDG 2022 \cite{PDG2022},
and display the SM Higgs potential, $|V_{eff}|^{1/4}$, in the left plot of Fig.~\ref{fig-nonmin}; the black solid (dashed) curve refers to positive (negative) values of the potential.
As can be seen, in this case one has $\phi_B \simeq 7.8\times 10^{16}$ GeV and $V_B^{1/4} \simeq  4.6 \times 10^{15}$ GeV (barely satisfying condition (\ref{HephiB})),
below the upper bound from CMB, $U_i^{1/4}<2.5 \times 10^{16}$ GeV.
We study how the potential gets modified for various values of the non-minimal coupling $\xi$.
The upper solid (red) curve is obtained by taking $\xi=\xi_{\rm min}=975$; metastability is indeed avoided, because the potential gets flattened at field values below $\phi_B$.
The other curves show the shape of the potential for other values of $\xi$. The short and long dashed curves are obtained by taking 
$\xi=\xi_{\rm min}-600$ and $\xi=\xi_{\rm min}-500$, respectively. 
In the latter case metastability is avoided, but the potential is not monotonously increasing; still it is stable and there is no unphysical deep minimum\footnote{We are disregarding the possibility of inflation with $\xi \lesssim \xi_{min}$, as it would require starting at the left of the potential barrier, where slow-roll conditions would not be naturally accounted for.
It would be worth to explore this possibility, that goes beyond the scope of the present paper, in future.}.
The lower dot-dashed curve is obtained by taking $\xi-\xi_{\rm min}=2000$, so that the flattening takes place at values slightly smaller than $\phi_B$,
whose value is indicated by the vertical grey line.

Notice that all these configurations, although displaying a plateau and stabilizing the Higgs potential, cannot work for metric Higgs inflation. As already mentioned, metric Higgs inflation would require $U_{i}^{1/4} =  (7.8-8.5) \times 10^{15}$ GeV, 
while here the plateau is at most as high as the barrier, $V_B^{1/4} \simeq  4.6 \times 10^{15}$ GeV.
This shows that a slightly smaller value of $m_t$ with respect to the one adopted in Fig.~\ref{fig-nonmin} might, instead, provide viable metric Higgs inflation for a configuration that would have been metastable in the pure SM, without the non-minimal coupling, as will be explained in Sec.~\ref{sec:metric}.

In the right plot of Fig.~\ref{fig-nonmin}, as another illustrative example, we fix $m_t=171.54$ GeV, case $P_{3d}$ in Tab.~\ref{table2}, as suggested in \cite{Garzelli:2023rvx}, and show the effect of varying $\xi$ around $\xi_{min}=6.4\times 10^8$.

\section{Higgs Inflation in the metric formulation}
\label{sec:metric}

In the framework of the metric formulation, here we study the various configurations of the Higgs potential flattened by a non-minimal coupling that allow for viable Higgs inflation.

\subsection{Slow-roll parameters for Higgs inflation in the metric formulation}

The slow-roll parameters are given by
\beq
\epsilon(t) = \frac{1}{2}  \left( \frac{d {U}({\bar \phi} (t))}{d{\bar \phi(t)} }  \frac{1}{{U}( {\bar \phi}(t))} \right)^2  \left(  \frac{ d {\bar \chi}(t) }{ d {\bar \phi}(t))} \right)^{-2}
\label{eq-eps}
\eeq
\beq
\eta(t)=   \frac{d^2 {U}({\bar \phi} (t))}{d{\bar \phi^2(t)} }  \frac{1}{{U}( {\bar \phi}(t))}   \left(  \frac{ d {\bar \chi}(t) }{ d {\bar \phi}(t))} \right)^{-2}  
-  \frac{d {U}({\bar \phi} (t))}{d{\bar \phi(t)} }  \frac{1}{{U}( {\bar \phi}(t))}   \left(  \frac{ d {\bar \chi}(t) }{ d {\bar \phi}(t))} \right)^{-3}  \frac{ d^2 {\bar \chi}(t) }{ d {\bar \phi}^2(t)} \,\,\, ,
\label{eq-eta}
\eeq
from which one can easily derive the following approximate expressions for the large field regime ($\xi \bar \phi^2 \gg 1$)
\beq
\epsilon(t) \approx \frac{4}{3} \frac{1}{\xi^2} \frac{1}{\bar \phi^4(t)}     \,\,\,\, , \,\,\,\,
\eta(t) \approx - \frac{4}{3} \frac{1}{\xi} \frac{1}{\bar \phi^2(t)} \,\, .
\eeq

Since inflation ends when $\epsilon \approx 1$, namely $\bar \phi_e^2 \approx \sqrt{\frac{4}{3}} \frac{1}{\xi}$,
we get
\beq
N = \frac{1}{\sqrt{2}} \int_{\bar \phi_i}^{\bar \phi_e} d\bar \phi \frac{1}{\epsilon^{1/2}} \frac{d \bar \chi}{d \bar \phi} 
\approx \frac{3}{4} \xi  {\bar \phi_i^2} \,\,\,   \Rightarrow \,\,\, \bar \phi_i^2 \approx \frac{4}{3} \frac{N}{\xi} \,\,\, .
\label{eq-Nefolds}
\eeq
At the beginning of inflation the slow-roll parameters are approximately given by
\beq
\epsilon_i \approx  \frac{3}{4}  \frac{1}{N^2} \,\,\, , \,\,\, \eta_i \approx - \frac{1}{N} \,\, ,\quad \textrm{with}\quad |\eta_i|\gg \epsilon_i\,,
\eeq
so that there are quite sharp predictions for the observable parameters
\beq
n_s \approx 1 - 6\epsilon_i + 2 \eta_i \approx  1- \frac{2}{N}  \,\,\,\, , \,\,\, r \approx 16 \epsilon_i \approx  \frac{12}{N^2}  \,\,\, .
\label{eq-ns-r}
\eeq
Taking $N=50-60$, the model prediction for $n_s$ is in the range $0.960-0.967$, which is well compatible with the cosmological data. 
Indeed, according to Ref.~\cite{Planck:2018jri}, observationally one has $n_s=0.9649 \pm 0.0042$ at 68\% CL.
Taking $N=50-60$, the model prediction for $r$ is the range $0.0048-0.0033$; this is below the present Planck limit,  $r_{0.002} < 0.056$ at 95\% CL \cite{Planck:2018jri},
and below the most stringent recent combined analysis of the BICEP, Keck Array and Planck data, which places an upper bound on the tensor-to-scalar ratio
$r < 0.036$ at the 95\% CL \cite{BICEP:2021xfz}.
 
As already mentioned, the amplitude of perturbations is
\beq
2.1 \times 10^{-9} \approx \Delta_R^2 \approx \frac{2}{3 \pi^2} \frac{\bar U_i}{r} \approx    \frac{1}{432 \pi^2} \lambda(t_{\rm lim}) \left( \frac{N}{\xi(t_{\rm lim}) }\right)^2 \,\, ,
\label{eq-approx}
\eeq
where the last approximation was obtained exploiting the approximation $r \approx 12/N^2$, and taking ${\bar U_i} \approx \bar U_{\rm lim}$.
In our analysis we find numerically the value of $\xi$ that gives the right value for the amplitude of perturbations, and then calculate the parameters $r$ and $n_s$. 
We checked that the right-hand side of the previous equation is a good approximation for such value of $\xi$.

\subsection{Metric Higgs inflation phenomenology for stable and metastable configurations}

We now study in more detail how the SM Higgs effective potential gets modified, by exploiting the flattening induced by the non-minimal coupling, to account for successful inflation.
The transition from small to large field values happens when $\bar \phi(t) \sim  \bar \phi_\xi = 1/\sqrt{\xi}$.
Let us study in turn the various scenarios for $m_t$, considering increasing values of the latter, going from stability to metastability as in Fig.~\ref{fig-shapes}, for the benchmark points shown in Tabs.~\ref{table1} and \ref{table2}.

\begin{itemize}

\item $m_t < m_t^i$.
 
This is the traditional approach of Higgs inflation, suggested in \cite{BezHiggs}. The idea is to  consider a stable configuration (without minima at high field values) for the Higgs potential,
and flatten it with a value of $\xi$ that correctly reproduces the observational data from CMB, in particular the amplitude of perturbations (while $n_s$ and $r$ do not depend on $\xi$). It is well known that such value is quite high, $\xi = \mathcal O(10^4)$, as derived in the pioneering tree-level calculation of \cite{BezHiggs}.
Our more sophisticated analysis indicates slightly smaller values, in particular for stable configurations not far from the inflection point configuration. 
Indeed, from Eq.~(\ref{eq-approx}) one realizes that for $\lambda \approx 10^{-2}$ (typical of a stable configuration, see Fig.~\ref{fig-lambda-all}), 
one needs $\xi \approx 3 \times 10^3$.

Let us now numerically inspect this situation, taking central values for $m_H$ and $\alpha_s^{(5)}$ and choosing two values for the top mass:
$m_t =170.4$ GeV, which is the lower $3\sigma$ value according to PDG 2022 \cite{PDG2022}, and $m_t =170.8$ GeV. 
These values correspond respectively to $\delta_t=-3.8 \times 10^{-3} $ and $\delta_t=-1.5 \times 10^{-3} $ in the notation of Eq.~(\ref{deltat}). 
We call $P_{1a}$ and $P_{1b}$ these configurations in Tab.~\ref{table1} and study them in the upper panels of Fig.~\ref{fig-P1}.
The solid (black) curve shows $|\bar V_{eff}|^{1/4}$, while the dashed (red) curve shows $\bar U^{1/4}$ for the indicated value of $\xi$, 
which accounts for the observed density perturbations, taking $N=60$. It turns out that the required value of $\xi$ is $\xi \approx 3000$ for $P_{1a}$, and goes down to $\xi \approx 2000$ for $P_{1b}$;
this is related to the fact that $\phi_\xi$ (shown in the upper horizontal axis) increases, when going from $P_{1a}$ to $P_{1b}$.
The dashed (pink) horizontal line represents the value of $\bar U_i^{1/4}$ required by metric Higgs inflation, that is the value obtained by taking $r=12/N^2=0.0031$ in Eq.~(\ref{eq-UiCMB}). 
The value $\bar \phi_\xi = 1/\sqrt{\xi}$, where the flattening starts to be effective, is shown in the upper horizontal axes.  
The vertical grey lines correspond to the beginning and the end of inflation, which are also emphasized with circles in top of the curve of $\bar U_i^{1/4}$.
Inflation thus takes place at field values slightly below the Planck scale, starting at $\bar \phi_i$ and ending at $\bar \phi_e$.
The predictions for $n_s$ and $r$, showed in the lower right corner of the plot, are consistent with those expected from the approximations in Eq.~(\ref{eq-ns-r})
and are perfectly compatible with observational data from CMB, as already discussed.

\begin{figure}[t!]
\begin{center}
\includegraphics[width=7.5cm]{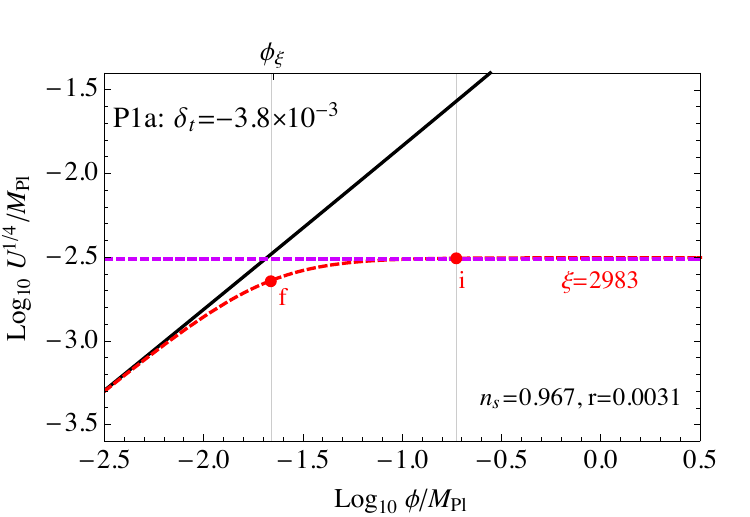}   \includegraphics[width=7.5cm]{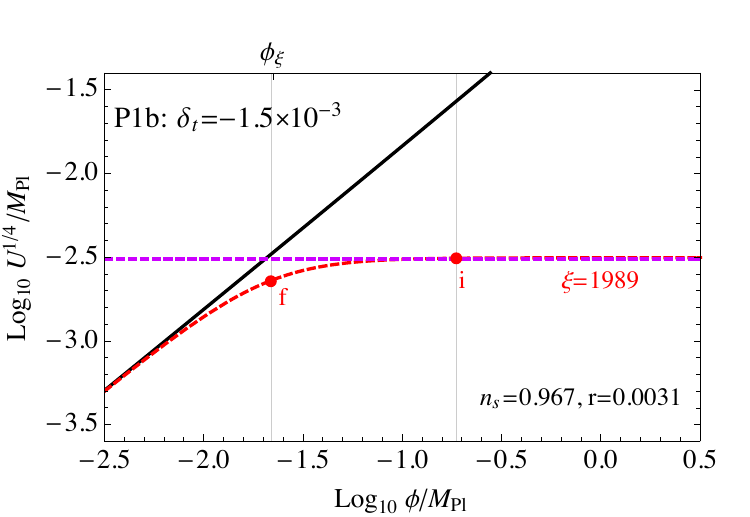}  \\   
\includegraphics[width=7.5cm]{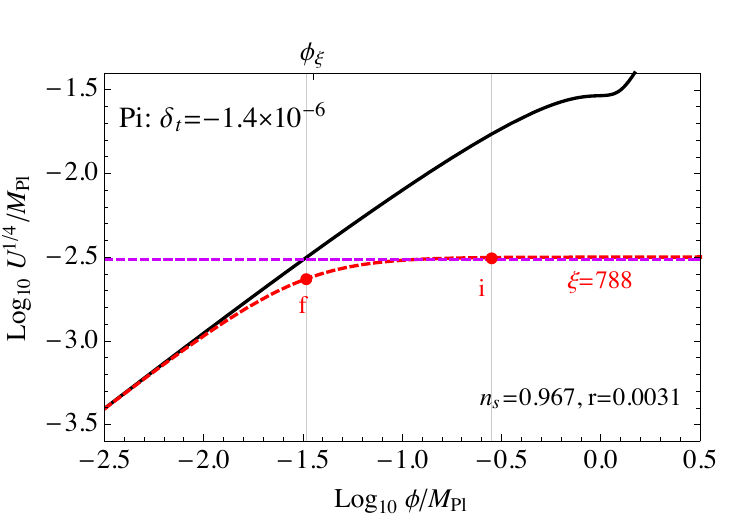}      \includegraphics[width=7.5cm]{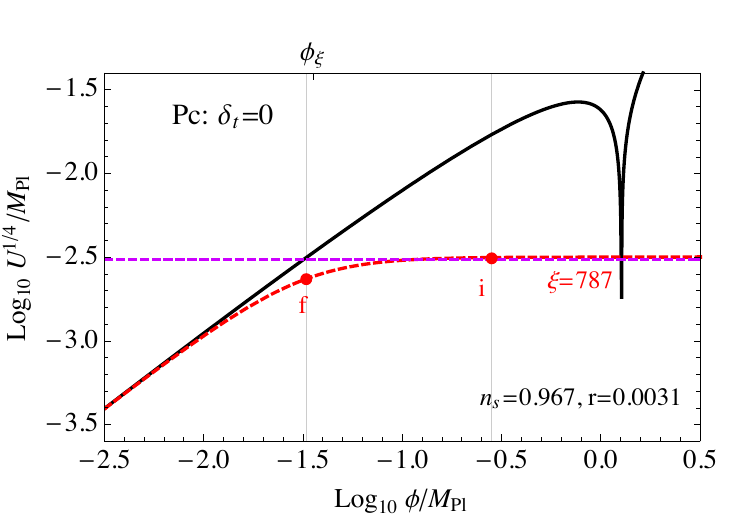}   \\
\includegraphics[width=7.5cm]{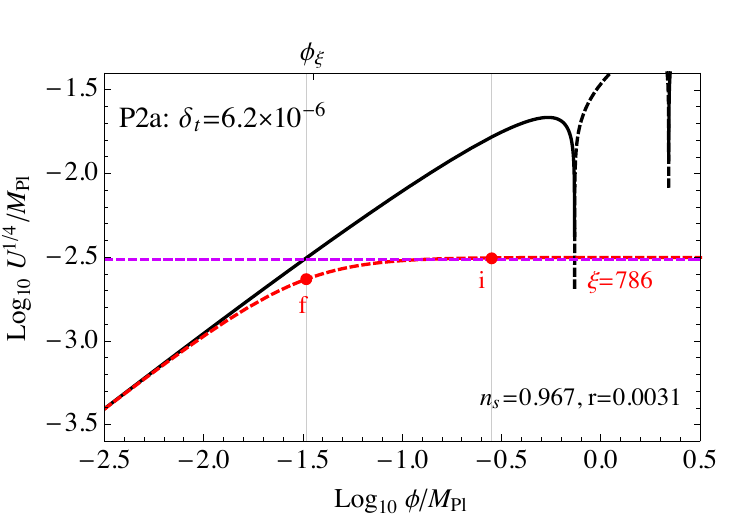}    \includegraphics[width=7.5cm]{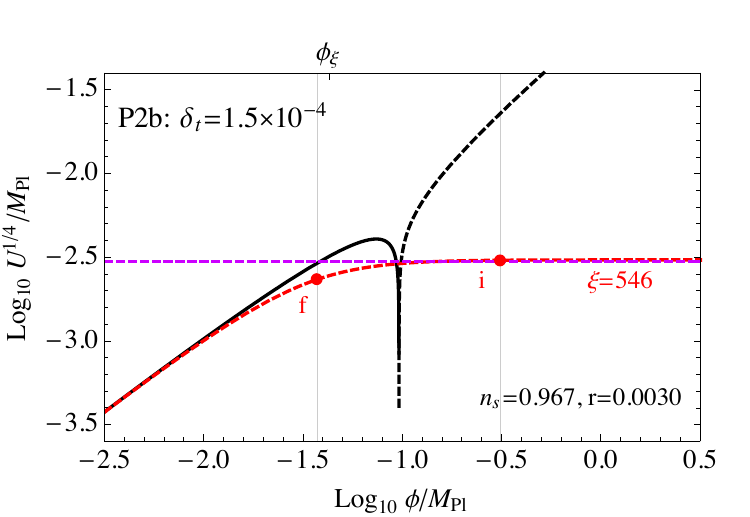}
\end{center}
\caption{\baselineskip=15 pt \small \it Various configurations, using central values for $m_H$ and $\alpha_s^{(5)}$. Increasing values of the top mass, as indicated.  
The shape of $|\bar V_{eff}|^{1/4}$ (solid black when positive, dashed black when negative) is shown, together with the shape of $\bar U^{1/4}$ (dashed red) 
for the indicated value of $\xi$, accounting for the observed density perturbations and taking $N=60$.}   
\label{fig-P1}
\vskip .5 cm
\end{figure}

\item $m_t = m_t^i$. 

Slightly rising $m_t$ with respect to $P_{1b}$, we find a configuration that corresponds to an inflection point, and which we call $P_i$ in Tab.~\ref{table1}, 
as displayed in the left middle panel of Fig.~\ref{fig-P1}. In this case, $\delta_t = -1.6 \times 10^{-6}$.
Essentially, one can envisage two scenarios which essentially depend on the value of the parameter $\xi$. 

\noindent i) The potential is flattened at field values larger than those of the inflection point, that is larger than $\phi_B$. 
This requires $\bar \phi_\xi >  \bar \phi_B \approx 2 \times 10^{18}$ GeV, as can be seen from Fig.\,\ref{fig-c}. 
Hence, one expects $\xi \lesssim\mathcal O(1)$, which might appear very interesting for inflation \cite{Bezcritical}; in addition, one might achieve a two-plateau potential~\footnote{It was suggested one could exploit the highest plateau induced by the non-minimal coupling at field values larger than $\phi_\xi$ for inflation,
and to exploit the inflection point (around 20-30 e-folds after the beginning of inflation) to trigger the production of a non-negligible amount of primordial black holes~\cite{Ezquiaga:2017fvi}. However, the detailed analysis of Ref. \cite{Bezrukov:2017dyv} showed that this is not possible.}. 
The inflationary plateau is at least as high as the inflection point of $V_{eff}$, since for field values up to $\phi_\xi$, and in particular around $\phi_B$, 
one has $U \approx V_{eff}$. This configuration however cannot account for the observational data from CMB~\cite{Masina:2018ejw}: 
numerically, $V_B^{1/4} \approx 6 \times 10^{16}$ GeV is larger than the CMB constraint, $U_{i}^{1/4} <  2.5 \times 10^{16}$ GeV, as can be seen from the left middle plot in Fig.~\ref{fig-P1}. 

\noindent ii) The potential is flattened at values smaller than $\phi_B$, that is $\bar \phi_\xi < \bar \phi_B$. 
In this case, the shape of the Higgs potential around the inflection point is drastically affected: the inflection point is simply not there anymore.
For the sake of Higgs inflation, this configuration is quite similar to the former cases, $P_{1a}$ and $P_{1b}$; 
notice however that in this case $\xi \approx 800$, which is a significantly smaller value with respect to the configurations $P_{1a}$ and $P_{1b}$. 
Indeed, $\lambda(t_{\rm lim})$ gets smaller increasing $m_t$, so that also $\xi \approx \xi(t_{\rm lim})$ gets smaller according to Eq.~(\ref{eq-approx}).
Notice also that inflation starts and ends at slightly larger field values with respect to $P_{1a}$ and $P_{1b}$.

\item $m_t^i< m_t \leq m_t^c$. 

This range corresponds to a configuration displaying a second high energy minimum close to the Planck scale, 
which can be shallow or quite deep but, in any case, not lower than the electroweak minimum.
All the arguments of the previous case $P_i$ apply here too. One thus needs $\bar \phi_\xi < \bar \phi_B$ and a further very mild reduction of $\xi$ is achieved.
We display in the right middle panel of Fig.~\ref{fig-P1} the critical configuration corresponding to two degenerate minima, denoted as $P_c$, and shown in Tab.~\ref{table1}. 
For the sake of inflation, essentially $P_c$ is equivalent to $P_i$.

\item $m_t> m_t^c$. 

For metastable configurations, the potential has to be flattened before the appearance of the deepest minimum at high energies,
so that again we need $\bar \phi_\xi < \bar \phi_B$. 
Increasing slightly $m_t$ with respect to $P_c$, we reach a configuration displaying a tiny metastability, which we denote by $P_{2a}$ in Tab.~\ref{table2}, such that $\delta_t=6.2 \times 10^{-6}$.
The barrier between the two minima is slightly smaller than it was for $P_c$, as can be seen in the left bottom panel of Fig.~\ref{fig-P1}; 
the solid (dashed) black line refers to positive (negative) values for $V_{eff}$. 
The value of $\xi$ allowing for successful Higgs inflation is thus negligibly smaller with respect to the case of $P_c$, and the cosmological observables are unchanged. 

Increasing further the top mass to the value $m_t=171.08$ GeV, corresponding to $\delta_t=1.5\times 10^{-4}$, 
we reach the most metastable configuration $P_{2b}$, for the case of metric Higgs inflation, in Tab.~\ref{table2}, shown in the right bottom panel of Fig.~\ref{fig-P1}.
Now the top of the barrier, $V_B^{1/4}$, is significantly reduced with respect to $P_{2a}$, getting closer to the value of the inflationary potential required by Higgs inflation, corresponding to the horizontal (pink) dashed line. 
The flattening of the potential, associated to $\bar \phi_\xi$, takes place just before the barrier; in this case $\bar \phi_\xi$ slightly moves to the right with respect to $P_{2a}$;
as an effect, we have a further reduction of the non-minimal coupling, $\xi \approx 550$.
Notice that inflation stars at field values where one would have had a negative potential in the SM. Also notice that by further increasing the value of $m_t$ one could go to lower values of the parameter $\xi$, although without reaching the potential value required by density perturbations, in which case one should introduce an extra scalar taking care of the CMB data, or else the Higgs, with a stable potential, is not a good candidate for inflaton. 

\end{itemize}

\begin{figure}[h!]
\vskip 1.cm 
 \begin{center}
\includegraphics[width=7.6cm]{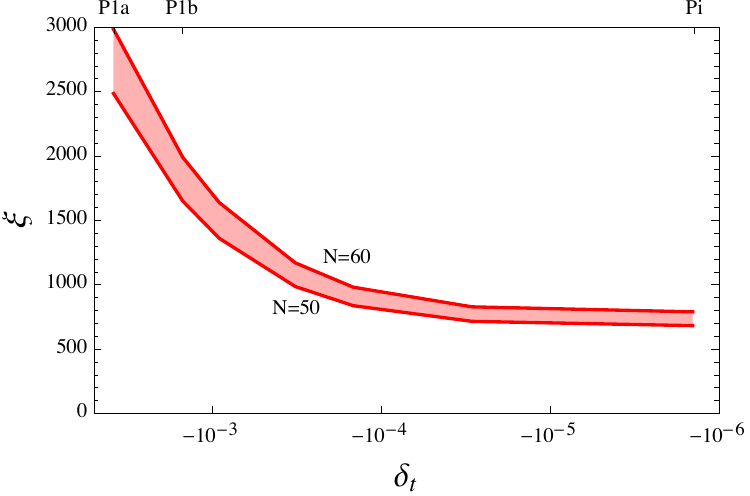}    \,\,\,  \hskip -.1 cm  \includegraphics[width=7.4cm]{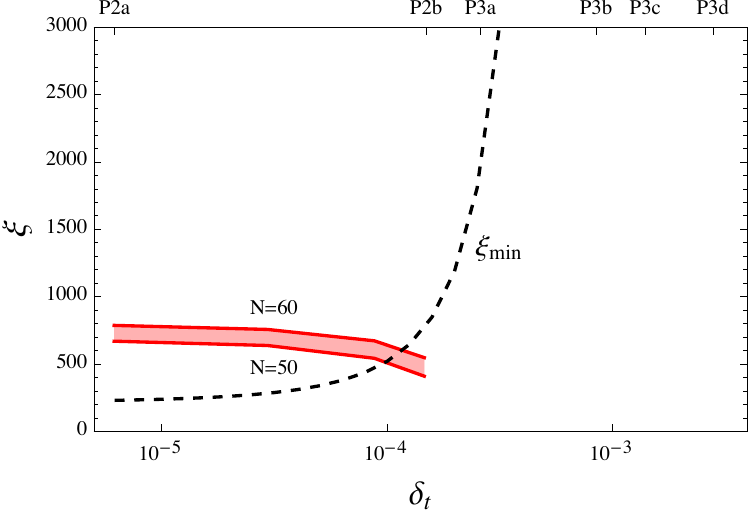}    
\end{center}
\caption{\baselineskip=15 pt \small  \it
The value of the non-minimal coupling in metric Higgs inflation, for stability (left panel) and metastability (right panel), for $N=50,\,60$ and taking the central values of the inputs, as in the PDG 2022~\cite{PDG2022}. For comparison, the dashed line $\xi_{\rm min}$ reproduces the values shown in the left panel of Fig.~\ref{fig-Barr}.}   
\label{fig-csi}
\vskip .5 cm
\end{figure}
We finally summarize the dependence of $\xi$, for metric Higgs inflation, on the top quark mass in Fig.~\ref{fig-csi} for values of the top mass corresponding to stability (left panel) and metastability (right panel). For genuinely stable configurations, we find the large value $\xi \sim 3000$, in agreement with the previous literature. In addition, notice that: 1) we are finding smaller $\xi$ values for the stable configurations that, in absence of the non-minimal coupling, would approach criticality, for which $\xi \sim 800$; 2) we can deal with cases where the Higgs potential (in the absence of the non-minimal coupling) is metastable, lowering $\xi$ down to about $500$. The right panel in Fig.~\ref{fig-csi} constitutes the main finding in this paper for the case of metric Higgs inflation. Moreover the dashed line in the right panel of Fig.~\ref{fig-csi} corresponds to the values of the parameter $\xi$ stabilizing the Higgs effective potential, without any relation with Higgs inflation.

\section{Higgs Inflation in the Palatini formulation}
\label{sec:Palatini}

In the framework of the Palatini formulation~\cite{Rasanen:2017ivk,Enckell:2018kkc,Shaposhnikov:2020fdv}, here we study the various configurations of the Higgs potential flattened by a non-minimal coupling that allow for viable Higgs inflation.

\subsection{Slow-roll parameters for Higgs inflation in the Palatini formulation}

From Eqs.~(\ref{eq-eps}) and (\ref{eq-eta}) and taking $\xi \bar \phi^2 \gg 1$, one derives the following approximate expressions for the slow-roll parameters in the Palatini formulation of Higgs inflation
\beq
\epsilon(t) \approx 8 \frac{1}{\xi} \frac{1}{\bar \phi^4(t)}     \,\,\,\, , \,\,\,\,
\eta(t) \approx - 8 \frac{1}{\bar \phi^2(t)} \,\, .
\eeq

In this case inflation ends when $\eta \approx 1$, namely $\bar \phi_e^2 \approx 8$,
and we get from Eq.~(\ref{eq-Nefolds}) 
\beq
N \approx \frac{1}{8}  {\bar \phi_i^2} \,\,\,   .
\eeq
At the beginning of inflation the slow-roll parameters are approximately given by
\beq
\epsilon_i \approx  \frac{1}{8}  \frac{1}{\xi N^2} \,\,\, , \,\,\, \eta_i \approx - \frac{1}{N} \,\, ,\quad \textrm{with}\quad |\eta_i|\gg \epsilon_i\,,
\eeq
so that the predictions for the observable parameters now become
\beq
n_s \approx 1 - 6\epsilon_i + 2 \eta_i \approx  1- \frac{2}{N}  \,\,\,\, , \,\,\, r \approx 16 \epsilon_i \approx  \frac{2}{\xi N^2}  \,\,\, .
\label{eq-ns-r-P}
\eeq
The prediction for $n_s$ is the same as in the metric formulation, while the prediction for $r$ is much more suppressed. Notice also that while $r$ is approximately constant in the metric formulation, in the Palatini one it is inversely proportional to $\xi$. 
 
The correct amplitude of density perturbations is now achieved with much smaller values of $\bar U_i$, according to
\beq
2.1 \times 10^{-9} \approx \Delta_R^2 \approx \frac{2}{3 \pi^2} \frac{\bar U_i}{r} 
\approx \frac{1}{ 72    \pi^2} { N^2}    \frac{\lambda(t_{\rm lim})}{ \xi(t_{\rm lim}) }
\label{eq-approx-P}
\eeq
where the last approximation was obtained exploiting the above approximation for $r$ and taking ${\bar U_i} \approx \bar U_{\rm lim}$.
We find numerically the value of $\xi$ that successfully reproduces the amplitude of perturbations, and then calculate the associated parameters $r$ and $n_s$.

\subsection{Palatini Higgs inflation phenomenology for stable and metastable configurations}

We now consider various scenarios with increasing values of $m_t$, going from stability to metastability, for some of the benchmark points shown in Tabs.~\ref{table1} and \ref{table2}.

\begin{figure}[htb]
\vskip 0.2cm 
\begin{center}
 \includegraphics[width=7.5cm]{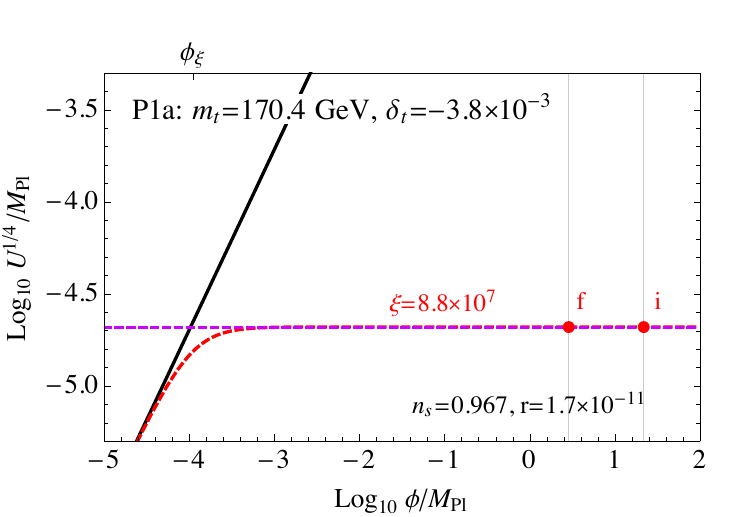}  \,\,\,, \,\,\, \includegraphics[width=7.5cm]{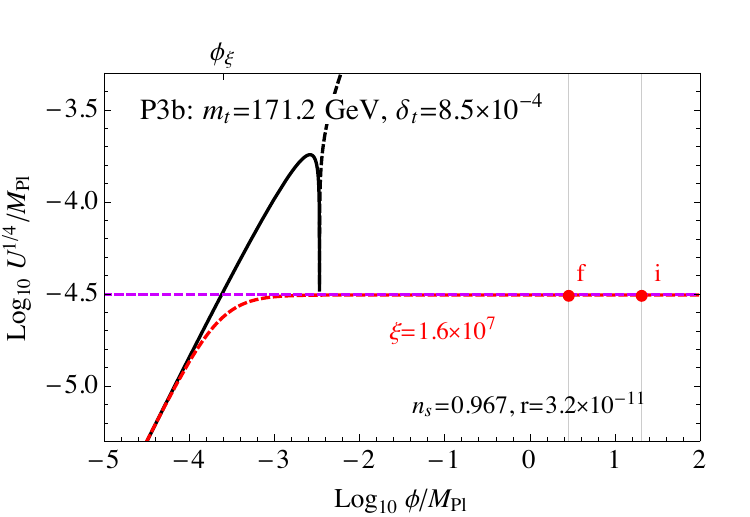}     \\ 
\includegraphics[width=7.5cm]{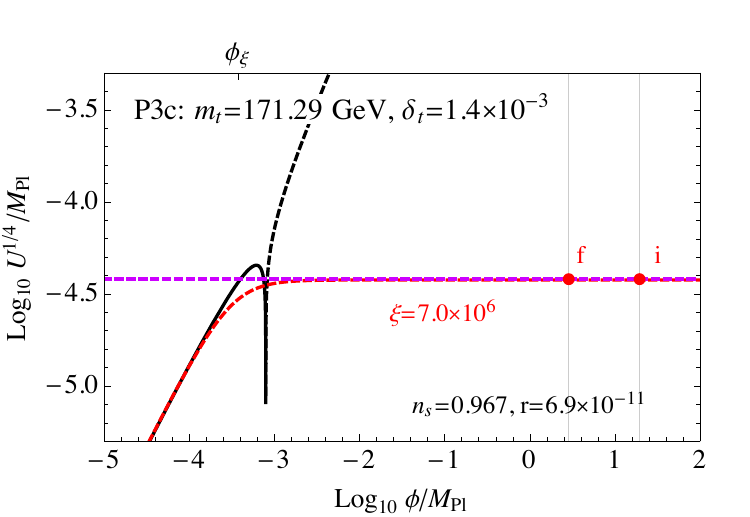}  \,\,\,, \,\,\, \includegraphics[width=7.5cm]{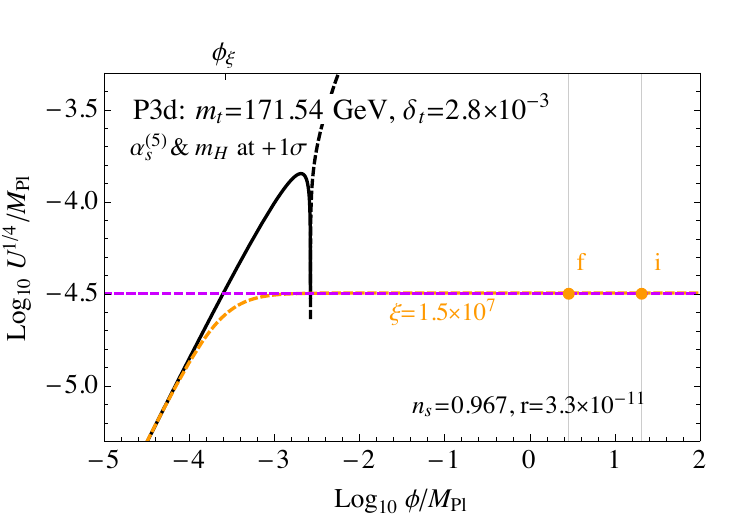}
 \end{center}
\caption{\baselineskip=15 pt \small \it 
Various configurations for the Palatini formulation. Central values for $m_H$ are adopted \cite{PDG2022}; the upper and bottom left plots are obtained taking $\alpha_s^{(5)}$ and $m_H$ at their central values, while for the bottom right $\alpha_s^{(5)}$ and $m_H$ are increased at their $1\sigma$ upper values~\cite{PDG2022}. The panels display increasing values of the top mass, as indicated.  
The shape of $|\bar V_{eff}|^{1/4}$ (solid black when positive, dashed black when negative) is shown, together with the shape of $\bar U^{1/4}$ (dashed red) 
for the indicated value of $\xi$, accounting for the observed density perturbations, assuming $N=60$.}   
\label{fig-P1-Pal}
\vskip .5 cm
\end{figure}

We first study, in the left upper panel of Fig.~\ref{fig-P1-Pal}, the benchmark point  $P_{1a}$ in Palatini formulation. The main difference with respect to the metric formulation is the much larger value needed for the non-minimal coupling, that is $\xi \approx 8.8 \times  10^{7}$. While nothing changes for $n_s$, the prediction for $r$ is significantly below the planned experimental sensitivity. Inflation takes place at values of $\phi$ larger than the Planck scale, and the corresponding value of the potential is $U^{1/4}_i \approx 5\times 10^{13}$ GeV.

These predictions are not significantly changed by increasing $m_t$ up to the value $171.2$ GeV, corresponding to $P_{3b}$, right upper panel of Fig.~\ref{fig-P1-Pal}; the main difference is a decrease in the required value of $\xi$.
It is possible to further increase the top mass up to about $171.29$ GeV, see $P_{3c}$, left lower panel in Fig.~\ref{fig-P1-Pal}, in which case $\xi \approx 7.0 \times 10^6$. Beyond the latter value of the top mass, the density of perturbations predicted by the model turns out to be too small with respect to observations.

Notice that the previous predictions are done using the central values of the parameters $\alpha_s^{(5)}$ and $m_H$. One can reach slightly larger values of $m_t$ by using the uncertainty in the determination of the electroweak parameters. For instance we show in the right lower panel of Fig.~\ref{fig-P1-Pal} the effective potential for the case $P_{3d}$ by fixing $\alpha_s^{(5)}$ and $m_H$ to the upper 1$\sigma$ limit of their experimental values, $\alpha_s^{(5)}=0.1188$ and $m_H=125.42$ GeV. This case corresponds to $\xi=1.5\times 10^7$ and $m_t=171.54$ GeV. The absolute upper bound we find for Palatini Higgs inflation is in this way $m_t=171.6$ GeV, which can be considered as our upper bound on the Higgs mass for Higgs inflation. Our results update and slightly improve over those found in Ref.~\cite{Shaposhnikov:2020fdv}.

\begin{figure}[h!]
\vskip .5cm 
 \begin{center}
\includegraphics[width=7.6cm]{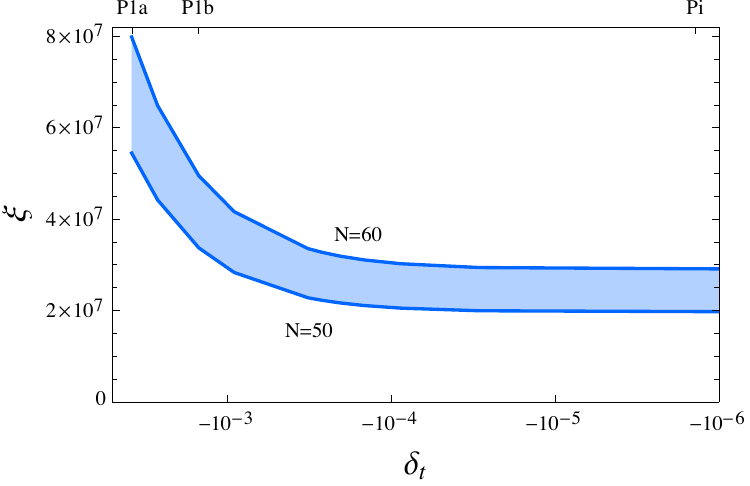}    \,\,\,  \hskip -.1 cm  \includegraphics[width=7.4cm]{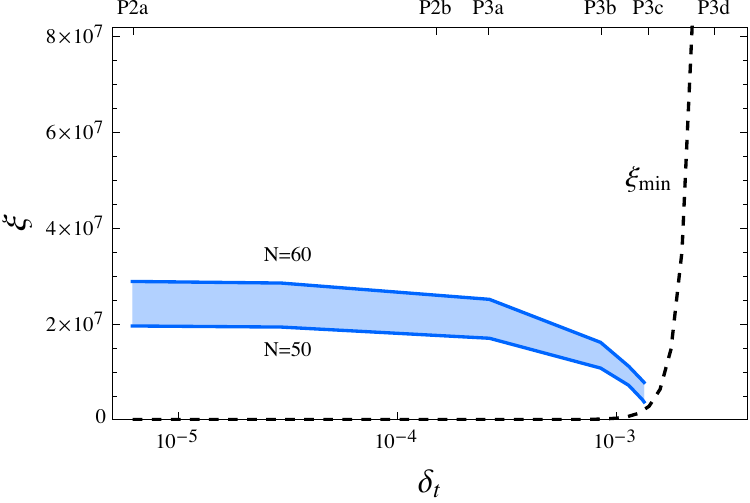}    
\end{center}
\caption{\baselineskip=15 pt \small  \it
The value of the non-minimal coupling in Palatini Higgs inflation, for stability (left panel) and metastability (right panel), for $N=50,\,60$ and taking the central values of the inputs, as in the PDG 2022~\cite{PDG2022}. For comparison, the dashed line $\xi_{\rm min}$ reproduces the values shown in the left panel of Fig.~\ref{fig-Barr}.}   
\label{fig-csiPal}
\vskip .5 cm
\end{figure}

As done for the case of metric Higgs inflation, we finally summarize the dependence of $\xi$, for Palatini Higgs inflation, on the top quark mass in Fig.~\ref{fig-csiPal} for values of the top mass corresponding to stability (left panel) and metastability (right panel). Notice that we can deal with cases where the Higgs potential is metastable so that the right panel in Fig.~\ref{fig-csiPal} constitutes the main finding in this paper for the case of Palatini Higgs inflation. Analogously the dashed line in the right panel of Fig.~\ref{fig-csiPal} corresponds to the values of the parameter $\xi$ stabilizing the Higgs effective potential, without any relation with Higgs inflation.

\section{Some comments on unitarity}
\label{sec:unitarity}

As we have seen earlier in this paper, imposing cosmological observables during the inflationary period, in particular the value of the CMB density fluctuations, requires a large value of the non-minimal Higgs coupling to gravity parameter $\xi\gg 1$. Such large value can be problematic because it may cause unitarity violation, as has been widely recognized in the literature~\cite{Burgess:2009ea,Barbon:2009ya,Burgess:2010zq,Hertzberg:2010dc,Bezrukov:2010jz,Lerner:2011it,Bezrukov:2013fka,Kehagias:2013mya,Ito:2021ssc,Antoniadis:2021axu}. In fact such large $\xi$ induces a cutoff $\Lambda$, lower than the Planck scale and can lead to violation of perturbative unitarity. 

The unitarity scale at large field values $\xi\phi^2\gg M_P^2$, during inflation, is $\Lambda_{1}\sim M_P/\sqrt{\xi}$\,.

As this cutoff is sufficiently larger than the Hubble scale $\Lambda_1\gtrsim H\sim M_P/\xi$, the inflationary dynamics is expected to be reliable. However, for metric Higgs inflation the cutoff at small field values $\xi\phi^2\ll M_P^2$ is $\Lambda_2\sim M_P/\xi$ \,,
so that reheating (involving inflaton oscillations) can be problematic, and usually requires UV completion to restore unitarity~\cite{Giudice:2010ka,Ema:2020zvg,Mikura:2021clt}. 
The problem of reheating in metric Higgs inflation has been studied in detail in Ref.~\cite{Sfakianakis:2018lzf} (but see also Refs. \cite{Bezrukov:2008ut, Garcia-Bellido:2008ycs, Ema:2016dny}) with the following result: 
\textit{i)} For $\xi\gtrsim 10^3$, instantaneous reheating takes place from gauge field modes production with maximum momentum $k_{\rm max}\gtrsim \Lambda_2$. This is associated to large reheating temperatures $T_{\rm reh}\sim k/3$, where $k$ is the typical momentum exchanged in particle scatterings inside the plasma. Therefore high momenta would violate the perturbative unitarity bound and the theory should be UV unitarized.
\textit{ii)} For $\xi\lesssim 10^3$ preheating proceeds through Higgs self-resonance production through scattering of Higgs bosons to the rest of the Standard Model fields. For such processes the preheating temperature is much lower and the modes momenta are $k\lesssim H\sim \Lambda_2$ thus, not violating unitarity in the reheating process, which is thus expected to be reliable~\cite{Sfakianakis:2018lzf}.  

As we can see from the plots of Fig.~\ref{fig-csi} for stable configurations in metric Higgs inflation cosmological observables require $\xi\gtrsim 10^3$ and high momenta should be involved in the reheating process violating the unitarity bound, while for metastable configurations we get $\xi<10^3$ suggesting that typical momenta in the reheating processes are below the unitarity bound.  

Finally, in the Palatini formulation of gravity it turns out that $\Lambda_1\sim \Lambda_2\sim M_P/\sqrt{\xi}$ so that Higgs inflation does not suffer from unitarity violation since the UV cutoff lies parametrically much higher than the Hubble rate~\cite{Bauer:2010jg}.
The unitarity issue can be then handled both in the inflationary and the reheating periods~\cite{Bauer:2010jg,Rubio:2019ypq,Mikura:2021clt,Yin:2022fgo}. Still the theory of Higgs inflation assumes that the inflationary and reheating periods will not be disrupted by the UV completion of the effective field theory. 

\section{Discussion and conclusions}
\label{sec:conclusions}

As cosmological inflation is supposed to be driven by a scalar field, and the only elementary scalar field in nature is the Higgs boson, identifying the inflaton with the Higgs boson has constituted a fascinating possibility, which is nevertheless in tension with CMB data on the amplitude of density perturbations. However, it was proven that such a problem can be solved if the Higgs is non-minimally coupled to gravity with a large coupling $\xi$, a theory dubbed Higgs inflation, in which case for large field values the Higgs potential is flattened and proportional to $\lambda(\mu)/\xi^2$, and one can cope with CMB data with a large enough value of $\xi$. Notice that this is just a tree-level flattening, as $\lambda$ runs with $\phi$ at loop level, according to the renormalization group equations. Thus this mechanism is jeopardized when including radiative corrections on the Higgs effective potential, as depending on the value of the quark top pole mass and the strong coupling constant, the effective potential, and the Higgs quartic coupling, become negative, detecting the existence of a deep minimum at high field values, and making the electroweak minimum metastable. This explains why Higgs inflation was originally proposed for stable configurations.

In this paper we have first reanalyzed the issue of the stability vs.~metastability of the Higgs effective potential using the most recent experimental data on the top quark and Higgs masses as well as the strong coupling, mainly in view of the capabilities of the Higgs to be identified with the inflaton triggering cosmological inflation and the seed of density perturbations in the early Universe. 

Further, we have exploited a particular feature of the Higgs effective potential in the Einstein frame, in cases where the Higgs is non-minimally coupled to gravity in the Jordan frame. In fact while for minimal coupling, $\xi=0$, perturbation theory in the effective potential is optimized for the choice of the renormalization scale $\bar\mu=\alpha \bar\phi$, for non minimal coupling it is optimized by the choice $\bar\mu=\alpha \bar\phi/\sqrt{1+\xi \bar\phi^2}$. This means that there is a limiting value of the renormalization scale given by $\bar\mu_{\rm\, lim}=\alpha /\sqrt{\xi}$ achieved for field values $\bar\phi>\bar\phi_\xi=1/\sqrt{\xi}$, and the effective potential is flattened to a value $\propto \lambda(\bar\mu_{\rm\, lim})/\xi^2$. Therefore a non-minimal coupling to gravity can naturally avoid a metastable minimum provided that $\xi\gtrsim \xi_{\rm min}$, where $\xi_{\rm min}\equiv 1/\bar\phi_B^2$.

Using the above tools we have (re)examined Higgs inflation, in the metric and Palatini formalisms of gravity, for the cases where:
\begin{description}
\item
\textit{i)} The electroweak vacuum is the only minimum of the effective potential (stability without barrier): the original proposal for Higgs inflation, valid for $m_t<m_t^i$.
\item
\textit{ii)} The electroweak vacuum is the true minimum, but there is either an inflection point or a second fake minimum, separated from the electroweak one by a potential barrier (stability with a barrier): valid for $m_t^i<m_t<m_t^c$.
\item
\textit{iii)} The electroweak vacuum is a fake minimum, separated by a barrier from the true non-physical AdS minimum (metastability): valid for $m_t>m_t^c$.

\end{description}

For a successful inflationary scenario two conditions are required:
\begin{description}
\item[a)]
The potential should be flattened for values of the field $\phi\lesssim \phi_B$, which requires a non-minimal coupling $\xi\gtrsim \xi_{\rm min}$. 
\item[b)]
The potential should reproduce the CMB condition on the amplitude of density perturbations.
\end{description}

We have then re-examined the different cases for the metric and Palatini formalisms. For the metric formalism, for the case \textit{i)} above, stability without barrier, condition a) is trivial as $\xi_{\rm min}=0$ so only condition b) applies, by which the flattening must take place about two orders of magnitude below the Planck scale, which yields values of $\xi\sim 3000$. For the case \textit{ii)}, with an inflection point or a barrier between the electroweak minimum and a local minimum, again $\xi_{\rm min}=0$ so that condition a) does not apply, while condition b) implies that the flattening of the potential must take place before the inflection point, or before the value of the field corresponding to the maximum of the barrier, respectively, leading to values $\xi\sim800 $. In particular for a value of the top quark mass equal to the critical value Higgs inflation can take place (critical Higgs inflation) for a value of the non minimal coupling $\xi=787$. Finally even metastable configurations, case \textit{iii)}, can satisfy conditions a) and b) above, and are viable for inflation. They require smaller values of $\xi$, down to $\xi \sim 500$. 
However, only metastable configurations close to critical configuration are allowed,
as the barrier between the two vacua must anyway be larger than the value of the inflaton potential suggested by CMB data for this model
(see the dashed purple lines in Fig.~\ref{fig-P1}).
This corresponds to just a tiny increase for $m_t$, $\sim 0.03$ GeV, with respect to $m_t^c$, corresponding to $\delta_ t \lesssim 1.5 \times 10^{-4}$ GeV.

For the Palatini formalism the qualitative description of the different cases is similar to that with the metric formalism, with two important differences: 
\textit{i)} The values of $\xi$ are much larger than those corresponding to the metric formalism. In fact they run from $\sim 10^8$ for the smallest values of $m_t$, to $\sim 10^7$ for the largest values of $m_t$. 
\textit{ii)} As a consequence of \textit{i)}, the values of the field where the potential can be flattened $\bar\phi_\xi=1/\sqrt{\xi}$ are smaller than in the metric formulation, thus increasing the values of the top quark mass for which Higgs inflation can be achieved. In fact we have found that for values up to $m_t\sim 171.6$ GeV Higgs inflation can take place.

In summary we have re-analyzed Higgs inflation in the presence of non-minimal coupling to gravity, beyond the stable configurations usually considered in the literature, both in the metric and in the Palatini formulations of gravity. We have proven that inflection and critical Higgs inflation are rescued as viable possibilities for suitable values of the non-minimal coupling. Moreover we have proven that even metastable configurations, going beyond the critical case, are viable possibilities for Higgs inflation, especially for the case of Palatini inflation.
This extension of the allowed range for $m_t$ is relevant in principle, as it opens up to metastable configurations. For larger values of the top quark mass cosmological inflation can take place but the predicted amplitude of density perturbations at CMB are too small, motivating the presence of a scalaron from the $R^2$ term in the Jordan frame. 

To conclude, considering the central values for the top quark pole mass and strong coupling constant quoted by the PDG 2022 data, the maximum value of the top quark mass consistent with Higgs inflation is 1.7$\sigma$ away from its central value. However taking into account the error in the strong coupling and the Higgs mass, the maximum allowed value deviates around 1.3$\sigma$ from the central experimental value of the top quark mass. Moreover, considering recent analyses~\cite{Garzelli:2023rvx} based on measurements of the top quark pole mass from differential cross sections by ATLAS and CMS, the deviation can shrink to within less than 1$\sigma$. So,
ultimately while Higgs inflation cannot be excluded by present data, its viability will depend on future experimental measurements of the top quark mass and the strong coupling constant.

\section*{\large Acknowledgments}

We thank the CERN Theory Department for kind hospitality and support during the completion of this work. 
One of us MQ would like to thank J.~R.~Espinosa for useful comments about the unitarity issue in Higgs inflation.
IM acknowledges partial support by the research project TAsP (Theoretical Astroparticle Physics) funded by the Istituto Nazionale di Fisica Nucleare (INFN). 
The work of MQ is supported by the Departament d'Empresa i Coneixement, Generalitat de Catalunya, Grant No.~2021 SGR 00649, and by the Ministerio de Econom\'ia y Competitividad, Grant No.~PID2020-115845GB-I00. IFAE is partially funded by the CERCA program of the Generalitat de Catalunya.

\appendix
\vskip 1.cm

\bibliographystyle{elsarticle-num} 
\bibliography{bib-meta} 

\begin{thebibliography}{10}
\expandafter\ifx\csname url\endcsname\relax
  \def\url#1{\texttt{#1}}\fi
\expandafter\ifx\csname urlprefix\endcsname\relax\def\urlprefix{URL }\fi
\expandafter\ifx\csname href\endcsname\relax
  \def\href#1#2{#2} \def\path#1{#1}\fi

\bibitem{Hung:1979dn}
P.~Q. Hung, {Vacuum Instability and New Constraints on Fermion Masses}, Phys.
  Rev. Lett. 42 (1979) 873.
\newblock \href {https://doi.org/10.1103/PhysRevLett.42.873}
  {\path{doi:10.1103/PhysRevLett.42.873}}.

\bibitem{Cabibbo:1979ay}
N.~Cabibbo, L.~Maiani, G.~Parisi, R.~Petronzio, {Bounds on the Fermions and
  Higgs Boson Masses in Grand Unified Theories}, Nucl. Phys. B158 (1979)
  295--305.
\newblock \href {https://doi.org/10.1016/0550-3213(79)90167-6}
  {\path{doi:10.1016/0550-3213(79)90167-6}}.

\bibitem{Froggatt:1995rt}
C.~D. Froggatt, H.~B. Nielsen, {Standard model criticality prediction: Top mass
  173 $\pm$ 5 GeV and Higgs mass 135 $\pm$ 9 GeV}, Phys. Lett. B368 (1996)
  96--102.
\newblock \href {http://arxiv.org/abs/hep-ph/9511371}
  {\path{arXiv:hep-ph/9511371}}, \href
  {https://doi.org/10.1016/0370-2693(95)01480-2}
  {\path{doi:10.1016/0370-2693(95)01480-2}}.

\bibitem{Degrassi}
G.~Degrassi, S.~Di~Vita, J.~Elias-Miro, J.~R. Espinosa, G.~F. Giudice, et~al.,
  {Higgs mass and vacuum stability in the Standard Model at NNLO}, JHEP 1208
  (2012) 098.
\newblock \href {http://arxiv.org/abs/1205.6497} {\path{arXiv:1205.6497}},
  \href {https://doi.org/10.1007/JHEP08(2012)098}
  {\path{doi:10.1007/JHEP08(2012)098}}.

\bibitem{Masina:2012tz}
I.~Masina, {Higgs boson and top quark masses as tests of electroweak vacuum
  stability}, Phys. Rev. D87~(5) (2013) 053001.
\newblock \href {http://arxiv.org/abs/1209.0393} {\path{arXiv:1209.0393}},
  \href {https://doi.org/10.1103/PhysRevD.87.053001}
  {\path{doi:10.1103/PhysRevD.87.053001}}.

\bibitem{Buttazzo:2013uya}
D.~Buttazzo, G.~Degrassi, P.~P. Giardino, G.~F. Giudice, F.~Sala, A.~Salvio,
  A.~Strumia, {Investigating the near-criticality of the Higgs boson}, JHEP 12
  (2013) 089.
\newblock \href {http://arxiv.org/abs/1307.3536} {\path{arXiv:1307.3536}},
  \href {https://doi.org/10.1007/JHEP12(2013)089}
  {\path{doi:10.1007/JHEP12(2013)089}}.

\bibitem{Bednyakov:2015sca}
A.~V. Bednyakov, B.~A. Kniehl, A.~F. Pikelner, O.~L. Veretin, {Stability of the
  Electroweak Vacuum: Gauge Independence and Advanced Precision}, Phys. Rev.
  Lett. 115~(20) (2015) 201802.
\newblock \href {http://arxiv.org/abs/1507.08833} {\path{arXiv:1507.08833}},
  \href {https://doi.org/10.1103/PhysRevLett.115.201802}
  {\path{doi:10.1103/PhysRevLett.115.201802}}.

\bibitem{Franceschini:2022veh}
R.~Franceschini, A.~Strumia, A.~Wulzer, {The collider landscape: which collider
  for establishing the SM instability?}, JHEP 08 (2022) 229, [Erratum: JHEP 03,
  167 (2023)].
\newblock \href {http://arxiv.org/abs/2203.17197} {\path{arXiv:2203.17197}},
  \href {https://doi.org/10.1007/JHEP08(2022)229}
  {\path{doi:10.1007/JHEP08(2022)229}}.

\bibitem{Hiller:2024zjp}
G.~Hiller, T.~H\"ohne, D.~F. Litim, T.~Steudtner, {Vacuum Stability in the
  Standard Model and Beyond} (1 2024).
\newblock \href {http://arxiv.org/abs/2401.08811} {\path{arXiv:2401.08811}}.

\bibitem{Iacobellis:2016eof}
G.~Iacobellis, I.~Masina, {Stationary configurations of the Standard Model
  Higgs potential: electroweak stability and rising inflection point}, Phys.
  Rev. D94~(7) (2016) 073005.
\newblock \href {http://arxiv.org/abs/1604.06046} {\path{arXiv:1604.06046}},
  \href {https://doi.org/10.1103/PhysRevD.94.073005}
  {\path{doi:10.1103/PhysRevD.94.073005}}.

\bibitem{Hamada:2014iga}
Y.~Hamada, H.~Kawai, K.-y. Oda, S.~C. Park, {Higgs Inflation is Still Alive
  after the Results from BICEP2}, Phys. Rev. Lett. 112~(24) (2014) 241301.
\newblock \href {http://arxiv.org/abs/1403.5043} {\path{arXiv:1403.5043}},
  \href {https://doi.org/10.1103/PhysRevLett.112.241301}
  {\path{doi:10.1103/PhysRevLett.112.241301}}.

\bibitem{Masina:2014yga}
I.~Masina, {The Gravitational wave background and Higgs false vacuum
  inflation}, Phys. Rev. D 89~(12) (2014) 123505.
\newblock \href {http://arxiv.org/abs/1403.5244} {\path{arXiv:1403.5244}},
  \href {https://doi.org/10.1103/PhysRevD.89.123505}
  {\path{doi:10.1103/PhysRevD.89.123505}}.

\bibitem{Bezcritical}
F.~Bezrukov, M.~Shaposhnikov, {Higgs inflation at the critical point}, Phys.
  Lett. B734 (2014) 249--254.
\newblock \href {http://arxiv.org/abs/1403.6078} {\path{arXiv:1403.6078}},
  \href {https://doi.org/10.1016/j.physletb.2014.05.074}
  {\path{doi:10.1016/j.physletb.2014.05.074}}.

\bibitem{Hamada:2014wna}
Y.~Hamada, H.~Kawai, K.-y. Oda, S.~C. Park, {Higgs inflation from Standard
  Model criticality}, Phys. Rev. D 91 (2015) 053008.
\newblock \href {http://arxiv.org/abs/1408.4864} {\path{arXiv:1408.4864}},
  \href {https://doi.org/10.1103/PhysRevD.91.053008}
  {\path{doi:10.1103/PhysRevD.91.053008}}.

\bibitem{Ezquiaga:2017fvi}
J.~M. Ezquiaga, J.~Garcia-Bellido, E.~Ruiz~Morales, {Primordial Black Hole
  production in Critical Higgs Inflation}, Phys. Lett. B776 (2018) 345--349.
\newblock \href {http://arxiv.org/abs/1705.04861} {\path{arXiv:1705.04861}},
  \href {https://doi.org/10.1016/j.physletb.2017.11.039}
  {\path{doi:10.1016/j.physletb.2017.11.039}}.

\bibitem{Salvio:2017oyf}
A.~Salvio, {Initial Conditions for Critical Higgs Inflation}, Phys. Lett. B 780
  (2018) 111--117.
\newblock \href {http://arxiv.org/abs/1712.04477} {\path{arXiv:1712.04477}},
  \href {https://doi.org/10.1016/j.physletb.2018.03.009}
  {\path{doi:10.1016/j.physletb.2018.03.009}}.

\bibitem{MasinaHiggsmass}
I.~Masina, A.~Notari, {The Higgs mass range from Standard Model false vacuum
  Inflation in scalar-tensor gravity}, Phys.Rev. D85 (2012) 123506.
\newblock \href {http://arxiv.org/abs/1112.2659} {\path{arXiv:1112.2659}},
  \href {https://doi.org/10.1103/PhysRevD.85.123506}
  {\path{doi:10.1103/PhysRevD.85.123506}}.

\bibitem{Masinatop}
I.~Masina, A.~Notari, {Standard Model False Vacuum Inflation: Correlating the
  Tensor-to-Scalar Ratio to the Top Quark and Higgs Boson masses},
  Phys.Rev.Lett. 108 (2012) 191302.
\newblock \href {http://arxiv.org/abs/1112.5430} {\path{arXiv:1112.5430}},
  \href {https://doi.org/10.1103/PhysRevLett.108.191302}
  {\path{doi:10.1103/PhysRevLett.108.191302}}.

\bibitem{Masinahybrid}
I.~Masina, A.~Notari, {Inflation from the Higgs field false vacuum with hybrid
  potential}, JCAP 1211 (2012) 031.
\newblock \href {http://arxiv.org/abs/1204.4155} {\path{arXiv:1204.4155}},
  \href {https://doi.org/10.1088/1475-7516/2012/11/031}
  {\path{doi:10.1088/1475-7516/2012/11/031}}.

\bibitem{Masina:2018ejw}
I.~Masina, {Ruling out Critical Higgs Inflation?}, Phys. Rev. D 98~(4) (2018)
  043536.
\newblock \href {http://arxiv.org/abs/1805.02160} {\path{arXiv:1805.02160}},
  \href {https://doi.org/10.1103/PhysRevD.98.043536}
  {\path{doi:10.1103/PhysRevD.98.043536}}.

\bibitem{Isidori:2007vm}
G.~Isidori, V.~S. Rychkov, A.~Strumia, N.~Tetradis, {Gravitational corrections
  to standard model vacuum decay}, Phys. Rev. D 77 (2008) 025034.
\newblock \href {http://arxiv.org/abs/0712.0242} {\path{arXiv:0712.0242}},
  \href {https://doi.org/10.1103/PhysRevD.77.025034}
  {\path{doi:10.1103/PhysRevD.77.025034}}.

\bibitem{Gialamas:2022gxv}
I.~D. Gialamas, A.~Karam, T.~D. Pappas, {Gravitational corrections to
  electroweak vacuum decay: metric vs. Palatini}, Phys. Lett. B 840 (2023)
  137885.
\newblock \href {http://arxiv.org/abs/2212.03052} {\path{arXiv:2212.03052}},
  \href {https://doi.org/10.1016/j.physletb.2023.137885}
  {\path{doi:10.1016/j.physletb.2023.137885}}.

\bibitem{Gialamas:2023emn}
I.~D. Gialamas, H.~Veerm\"ae, {Electroweak vacuum decay in metric-affine
  gravity}, Phys. Lett. B 844 (2023) 138109.
\newblock \href {http://arxiv.org/abs/2305.07693} {\path{arXiv:2305.07693}},
  \href {https://doi.org/10.1016/j.physletb.2023.138109}
  {\path{doi:10.1016/j.physletb.2023.138109}}.

\bibitem{Salvio:2018rv}
A.~Salvio, {Critical Higgs inflation in a Viable Motivated Model}, Phys. Rev. D
  99~(1) (2019) 015037.
\newblock \href {http://arxiv.org/abs/1810.00792} {\path{arXiv:1810.00792}},
  \href {https://doi.org/10.1103/PhysRevD.99.015037}
  {\path{doi:10.1103/PhysRevD.99.015037}}.

\bibitem{Hamada:2013mya}
Y.~Hamada, H.~Kawai, K.-y. Oda, {Minimal Higgs inflation}, PTEP 2014 (2014)
  023B02.
\newblock \href {http://arxiv.org/abs/1308.6651} {\path{arXiv:1308.6651}},
  \href {https://doi.org/10.1093/ptep/ptt116} {\path{doi:10.1093/ptep/ptt116}}.

\bibitem{BezHiggs}
F.~Bezrukov, M.~Shaposhnikov, {The Standard Model Higgs boson as the inflaton},
  Phys.Lett. B659 (2008) 703--706.
\newblock \href {http://arxiv.org/abs/0710.3755} {\path{arXiv:0710.3755}},
  \href {https://doi.org/10.1016/j.physletb.2007.11.072}
  {\path{doi:10.1016/j.physletb.2007.11.072}}.

\bibitem{Barvinsky}
A.~Barvinsky, A.~Y. Kamenshchik, A.~Starobinsky, {Inflation scenario via the
  Standard Model Higgs boson and LHC}, JCAP 0811 (2008) 021.
\newblock \href {http://arxiv.org/abs/0809.2104} {\path{arXiv:0809.2104}},
  \href {https://doi.org/10.1088/1475-7516/2008/11/021}
  {\path{doi:10.1088/1475-7516/2008/11/021}}.

\bibitem{DeSimone:2008ei}
A.~De~Simone, M.~P. Hertzberg, F.~Wilczek, {Running Inflation in the Standard
  Model}, Phys. Lett. B678 (2009) 1--8.
\newblock \href {http://arxiv.org/abs/0812.4946} {\path{arXiv:0812.4946}},
  \href {https://doi.org/10.1016/j.physletb.2009.05.054}
  {\path{doi:10.1016/j.physletb.2009.05.054}}.

\bibitem{Beztwoloop}
F.~Bezrukov, M.~Shaposhnikov, {Standard Model Higgs boson mass from inflation:
  Two loop analysis}, JHEP 0907 (2009) 089.
\newblock \href {http://arxiv.org/abs/0904.1537} {\path{arXiv:0904.1537}},
  \href {https://doi.org/10.1088/1126-6708/2009/07/089}
  {\path{doi:10.1088/1126-6708/2009/07/089}}.

\bibitem{Barvinskyrenorm}
A.~Barvinsky, A.~Y. Kamenshchik, C.~Kiefer, A.~Starobinsky, C.~Steinwachs,
  {Higgs boson, renormalization group, and naturalness in cosmology},
  Eur.Phys.J. C72 (2012) 2219.
\newblock \href {http://arxiv.org/abs/0910.1041} {\path{arXiv:0910.1041}},
  \href {https://doi.org/10.1140/epjc/s10052-012-2219-3}
  {\path{doi:10.1140/epjc/s10052-012-2219-3}}.

\bibitem{Bezrukov:2010jz}
F.~Bezrukov, A.~Magnin, M.~Shaposhnikov, S.~Sibiryakov, {Higgs inflation:
  consistency and generalisations}, JHEP 01 (2011) 016.
\newblock \href {http://arxiv.org/abs/1008.5157} {\path{arXiv:1008.5157}},
  \href {https://doi.org/10.1007/JHEP01(2011)016}
  {\path{doi:10.1007/JHEP01(2011)016}}.

\bibitem{Bezrukov:2012sa}
F.~Bezrukov, M.~{\relax Yu}. Kalmykov, B.~A. Kniehl, M.~Shaposhnikov, {Higgs
  Boson Mass and New Physics}, JHEP 10 (2012) 140.
\newblock \href {http://arxiv.org/abs/1205.2893} {\path{arXiv:1205.2893}},
  \href {https://doi.org/10.1007/JHEP10(2012)140}
  {\path{doi:10.1007/JHEP10(2012)140}}.

\bibitem{Allison}
K.~Allison, {Higgs xi-inflation for the 125-126 GeV Higgs: a two-loop
  analysis}, JHEP 1402 (2014) 040.
\newblock \href {http://arxiv.org/abs/1306.6931} {\path{arXiv:1306.6931}},
  \href {https://doi.org/10.1007/JHEP02(2014)040}
  {\path{doi:10.1007/JHEP02(2014)040}}.

\bibitem{Bezrukov:2014ina}
F.~Bezrukov, M.~Shaposhnikov, {Why should we care about the top quark Yukawa
  coupling?}, J. Exp. Theor. Phys. 120 (2015) 335--343, [Zh. Eksp. Teor.
  Fiz.147,389(2015)].
\newblock \href {http://arxiv.org/abs/1411.1923} {\path{arXiv:1411.1923}},
  \href {https://doi.org/10.1134/S1063776115030152}
  {\path{doi:10.1134/S1063776115030152}}.

\bibitem{Bezrukov:2017dyv}
F.~Bezrukov, M.~Pauly, J.~Rubio, {On the robustness of the primordial power
  spectrum in renormalized Higgs inflation}, JCAP 1802~(02) (2018) 040.
\newblock \href {http://arxiv.org/abs/1706.05007} {\path{arXiv:1706.05007}},
  \href {https://doi.org/10.1088/1475-7516/2018/02/040}
  {\path{doi:10.1088/1475-7516/2018/02/040}}.

\bibitem{PDG2022}
R.~L. Workman, et~al., The review of particle physics, 2022 and 2023 update,
  Prog. Theor. Exp. Phys. 083C01 (2022).

\bibitem{Bezrukov:2014ipa}
F.~Bezrukov, J.~Rubio, M.~Shaposhnikov, {Living beyond the edge: Higgs
  inflation and vacuum metastability}, Phys. Rev. D 92~(8) (2015) 083512.
\newblock \href {http://arxiv.org/abs/1412.3811} {\path{arXiv:1412.3811}},
  \href {https://doi.org/10.1103/PhysRevD.92.083512}
  {\path{doi:10.1103/PhysRevD.92.083512}}.

\bibitem{Ford:1992}
C.~Ford, I.~Jack, D.~R.~T. Jones, {The Standard model effective potential at
  two loops}, Nucl. Phys. B387 (1992) 373--390, [Erratum: Nucl.
  Phys.B504,551(1997)].
\newblock \href {http://arxiv.org/abs/hep-ph/0111190}
  {\path{arXiv:hep-ph/0111190}}, \href
  {https://doi.org/10.1016/0550-3213(92)90165-8}
  {\path{doi:10.1016/0550-3213(92)90165-8}}.

\bibitem{Coleman}
S.~Coleman, E.~Weinberg, {Radiative Corrections as the Origin of Spontaneous
  Symmetry Breaking}, Phys. Rev. D 7 (1973) 1888--1910.
\newblock \href {https://doi.org/10.1103/PhysRevD.7.1888}
  {\path{doi:10.1103/PhysRevD.7.1888}}.

\bibitem{Alam:2022cdv}
Z.~Alam, S.~P. Martin, {Standard model at 200~GeV}, Phys. Rev. D 107~(1) (2023)
  013010.
\newblock \href {http://arxiv.org/abs/2211.08576} {\path{arXiv:2211.08576}},
  \href {https://doi.org/10.1103/PhysRevD.107.013010}
  {\path{doi:10.1103/PhysRevD.107.013010}}.

\bibitem{CMS:2023wnd}
{Combination of measurements of the top quark mass from data collected by the
  ATLAS and CMS experiments at $\sqrt{s}=7$ and $8~\mathrm{TeV}$} (2023).

\bibitem{Garzelli:2023rvx}
M.~V. Garzelli, J.~Mazzitelli, S.~O. Moch, O.~Zenaiev, {Top-quark pole mass
  extraction at NNLO accuracy, from total, single- and double-differential
  cross sections for $t\bar{t}+X$ production at the LHC} (11 2023).
\newblock \href {http://arxiv.org/abs/2311.05509} {\path{arXiv:2311.05509}}.

\bibitem{Mihaila:2012}
L.~N. Mihaila, J.~Salomon, M.~Steinhauser, {Gauge Coupling $\beta$-functions in
  the Standard Model to Three Loops}, Phys. Rev. Lett. 108 (2012) 151602.
\newblock \href {http://arxiv.org/abs/1201.5868} {\path{arXiv:1201.5868}},
  \href {https://doi.org/10.1103/PhysRevLett.108.151602}
  {\path{doi:10.1103/PhysRevLett.108.151602}}.

\bibitem{Mihaila1}
L.~N. Mihaila, J.~Salomon, M.~Steinhauser, {Renormalization constants and
  $\beta$-functions for the gauge couplings of the Standard Model to three-loop
  order}, Phys. Rev. D86 (2012) 096008.
\newblock \href {http://arxiv.org/abs/1208.3357} {\path{arXiv:1208.3357}},
  \href {https://doi.org/10.1103/PhysRevD.86.096008}
  {\path{doi:10.1103/PhysRevD.86.096008}}.

\bibitem{ChetyrkinZoller}
K.~G. Chetyrkin, M.~F. Zoller, {Three-loop $\beta$-functions for top-Yukawa and
  the Higgs self-interaction in the Standard Model}, JHEP 06 (2012) 033.
\newblock \href {http://arxiv.org/abs/1205.2892} {\path{arXiv:1205.2892}},
  \href {https://doi.org/10.1007/JHEP06(2012)033}
  {\path{doi:10.1007/JHEP06(2012)033}}.

\bibitem{Chetyrkin:2013}
K.~G. Chetyrkin, M.~F. Zoller, {$\beta$-function for the Higgs self-interaction
  in the Standard Model at three-loop level}, JHEP 04 (2013) 091, [Erratum:
  JHEP09,155(2013)].
\newblock \href {http://arxiv.org/abs/1303.2890} {\path{arXiv:1303.2890}},
  \href {https://doi.org/10.1007/JHEP04(2013)091, 10.1007/JHEP09(2013)155}
  {\path{doi:10.1007/JHEP04(2013)091, 10.1007/JHEP09(2013)155}}.

\bibitem{BednyakovPikelnerVelizhanin}
A.~V. Bednyakov, A.~F. Pikelner, V.~N. Velizhanin, {Higgs self-coupling
  $\beta$-function in the Standard Model at three loops}, Nucl. Phys. B875
  (2013) 552--565.
\newblock \href {http://arxiv.org/abs/1303.4364} {\path{arXiv:1303.4364}},
  \href {https://doi.org/10.1016/j.nuclphysb.2013.07.015}
  {\path{doi:10.1016/j.nuclphysb.2013.07.015}}.

\bibitem{BednyakovPikelnerVelizhanin1}
A.~V. Bednyakov, A.~F. Pikelner, V.~N. Velizhanin, {Yukawa coupling
  $\beta$-functions in the Standard Model at three loops}, Phys. Lett. B722
  (2013) 336--340.
\newblock \href {http://arxiv.org/abs/1212.6829} {\path{arXiv:1212.6829}},
  \href {https://doi.org/10.1016/j.physletb.2013.04.038}
  {\path{doi:10.1016/j.physletb.2013.04.038}}.

\bibitem{Bednyakov:2013}
A.~V. Bednyakov, A.~F. Pikelner, V.~N. Velizhanin, {Three-loop Higgs
  self-coupling $\beta$-function in the Standard Model with complex Yukawa
  matrices}, Nucl. Phys. B879 (2014) 256--267.
\newblock \href {http://arxiv.org/abs/1310.3806} {\path{arXiv:1310.3806}},
  \href {https://doi.org/10.1016/j.nuclphysb.2013.12.012}
  {\path{doi:10.1016/j.nuclphysb.2013.12.012}}.

\bibitem{Bednyakov:2014}
A.~V. Bednyakov, A.~F. Pikelner, V.~N. Velizhanin, {Three-loop SM
  $\beta$-functions for matrix Yukawa couplings}, Phys. Lett. B737 (2014)
  129--134.
\newblock \href {http://arxiv.org/abs/1406.7171} {\path{arXiv:1406.7171}},
  \href {https://doi.org/10.1016/j.physletb.2014.08.049}
  {\path{doi:10.1016/j.physletb.2014.08.049}}.

\bibitem{Zoller:2015tha}
M.~F. Zoller, {Top-Yukawa effects on the $\beta$-function of the strong
  coupling in the SM at four-loop level}, JHEP 02 (2016) 095.
\newblock \href {http://arxiv.org/abs/1508.03624} {\path{arXiv:1508.03624}},
  \href {https://doi.org/10.1007/JHEP02(2016)095}
  {\path{doi:10.1007/JHEP02(2016)095}}.

\bibitem{Bednyakov:2015}
A.~V. Bednyakov, A.~F. Pikelner, {Four-loop strong coupling $\beta$-function in
  the Standard Model. }, Physics Letters B 762 (2016) 151--156.
\newblock \href {http://arxiv.org/abs/1508.02680} {\path{arXiv:1508.02680}}.

\bibitem{Chetyrkin:2016ruf}
K.~G. Chetyrkin, M.~F. Zoller, {Leading QCD-induced four-loop contributions to
  the $\beta$-function of the Higgs self-coupling in the SM and vacuum
  stability.}, J. High Energ. Phys. 175 (2016).
\newblock \href {http://arxiv.org/abs/1604.00853} {\path{arXiv:1604.00853}}.

\bibitem{Dav:2020}
J.~Davies, F.~Herren, C.~Poole, M.~Steinhauser, E.~Thomsen, A.\, Gauge coupling
  $\ensuremath{\beta}$ functions to four-loop order in the standard model,
  Phys. Rev. Lett. 124 (2020) 071803.

\bibitem{Coleman:1973jx}
S.~R. Coleman, E.~J. Weinberg, {Radiative Corrections as the Origin of
  Spontaneous Symmetry Breaking}, Phys. Rev. D7 (1973) 1888--1910.
\newblock \href {https://doi.org/10.1103/PhysRevD.7.1888}
  {\path{doi:10.1103/PhysRevD.7.1888}}.

\bibitem{Martin:2017lqn}
S.~P. Martin, {Effective potential at three loops}, Phys. Rev. D 96~(9) (2017)
  096005.
\newblock \href {http://arxiv.org/abs/1709.02397} {\path{arXiv:1709.02397}},
  \href {https://doi.org/10.1103/PhysRevD.96.096005}
  {\path{doi:10.1103/PhysRevD.96.096005}}.

\bibitem{Martin:2015eia}
S.~P. Martin, {Four-Loop Standard Model Effective Potential at Leading Order in
  QCD}, Phys. Rev. D 92~(5) (2015) 054029.
\newblock \href {http://arxiv.org/abs/1508.00912} {\path{arXiv:1508.00912}},
  \href {https://doi.org/10.1103/PhysRevD.92.054029}
  {\path{doi:10.1103/PhysRevD.92.054029}}.

\bibitem{Andreassen:2014gha}
A.~Andreassen, W.~Frost, M.~D. Schwartz, {Consistent Use of the Standard Model
  Effective Potential}, Phys. Rev. Lett. 113~(24) (2014) 241801.
\newblock \href {http://arxiv.org/abs/1408.0292} {\path{arXiv:1408.0292}},
  \href {https://doi.org/10.1103/PhysRevLett.113.241801}
  {\path{doi:10.1103/PhysRevLett.113.241801}}.

\bibitem{Nielsen:1975fs}
N.~K. Nielsen, {On the Gauge Dependence of Spontaneous Symmetry Breaking in
  Gauge Theories}, Nucl. Phys. B101 (1975) 173.
\newblock \href {https://doi.org/10.1016/0550-3213(75)90301-6}
  {\path{doi:10.1016/0550-3213(75)90301-6}}.

\bibitem{Casas:1996aq}
J.~A. Casas, J.~R. Espinosa, M.~Quiros, {Standard model stability bounds for
  new physics within LHC reach}, Phys. Lett. B382 (1996) 374--382.
\newblock \href {http://arxiv.org/abs/hep-ph/9603227}
  {\path{arXiv:hep-ph/9603227}}, \href
  {https://doi.org/10.1016/0370-2693(96)00682-X}
  {\path{doi:10.1016/0370-2693(96)00682-X}}.

\bibitem{BICEP:2021xfz}
P.~A.~R. Ade, et~al., {Improved Constraints on Primordial Gravitational Waves
  using Planck, WMAP, and BICEP/Keck Observations through the 2018 Observing
  Season}, Phys. Rev. Lett. 127~(15) (2021) 151301.
\newblock \href {http://arxiv.org/abs/2110.00483} {\path{arXiv:2110.00483}},
  \href {https://doi.org/10.1103/PhysRevLett.127.151301}
  {\path{doi:10.1103/PhysRevLett.127.151301}}.

\bibitem{EliasMiro:2012ay}
J.~Elias-Miro, J.~R. Espinosa, G.~F. Giudice, H.~M. Lee, A.~Strumia,
  {Stabilization of the Electroweak Vacuum by a Scalar Threshold Effect}, JHEP
  06 (2012) 031.
\newblock \href {http://arxiv.org/abs/1203.0237} {\path{arXiv:1203.0237}},
  \href {https://doi.org/10.1007/JHEP06(2012)031}
  {\path{doi:10.1007/JHEP06(2012)031}}.

\bibitem{Cado:2022evn}
Y.~Cado, M.~Quir\'os, {Baryogenesis from combined Higgs\textendash{}scalar
  field inflation}, Phys. Rev. D 106~(5) (2022) 055018.
\newblock \href {http://arxiv.org/abs/2201.06422} {\path{arXiv:2201.06422}},
  \href {https://doi.org/10.1103/PhysRevD.106.055018}
  {\path{doi:10.1103/PhysRevD.106.055018}}.

\bibitem{Cado:2023zbm}
Y.~Cado, C.~Englert, T.~Modak, M.~Quir\'os, {Baryogenesis in $R^2$-Higgs
  Inflation: the Gravitational Connection, } (12 2023).
\newblock \href {http://arxiv.org/abs/2312.10414} {\path{arXiv:2312.10414}}.

\bibitem{Yin:2022fgo}
W.~Yin, {Weak-Scale Higgs Inflation, } (10 2022).
\newblock \href {http://arxiv.org/abs/2210.15680} {\path{arXiv:2210.15680}}.

\bibitem{Palatini:1919ffw}
A.~Palatini, {Deduzione invariantiva delle equazioni gravitazionali dal
  principio di Hamilton}, Rend. Circ. Mat. Palermo 43~(1) (1919) 203--212.
\newblock \href {https://doi.org/10.1007/BF03014670}
  {\path{doi:10.1007/BF03014670}}.

\bibitem{George:2015nza}
D.~P. George, S.~Mooij, M.~Postma, {Quantum corrections in Higgs inflation: the
  Standard Model case}, JCAP 1604~(04) (2016) 006.
\newblock \href {http://arxiv.org/abs/1508.04660} {\path{arXiv:1508.04660}},
  \href {https://doi.org/10.1088/1475-7516/2016/04/006}
  {\path{doi:10.1088/1475-7516/2016/04/006}}.

\bibitem{Fumagalli:2016lls}
J.~Fumagalli, M.~Postma, {UV (in)sensitivity of Higgs inflation}, JHEP 05
  (2016) 049.
\newblock \href {http://arxiv.org/abs/1602.07234} {\path{arXiv:1602.07234}},
  \href {https://doi.org/10.1007/JHEP05(2016)049}
  {\path{doi:10.1007/JHEP05(2016)049}}.

\bibitem{Espinosa:2015qea}
J.~R. Espinosa, G.~F. Giudice, E.~Morgante, A.~Riotto, L.~Senatore, A.~Strumia,
  N.~Tetradis, {The cosmological Higgstory of the vacuum instability}, JHEP 09
  (2015) 174.
\newblock \href {http://arxiv.org/abs/1505.04825} {\path{arXiv:1505.04825}},
  \href {https://doi.org/10.1007/JHEP09(2015)174}
  {\path{doi:10.1007/JHEP09(2015)174}}.

\bibitem{Planck:2018jri}
Y.~Akrami, et~al., {Planck 2018 results. X. Constraints on inflation}, Astron.
  Astrophys. 641 (2020) A10.
\newblock \href {http://arxiv.org/abs/1807.06211} {\path{arXiv:1807.06211}},
  \href {https://doi.org/10.1051/0004-6361/201833887}
  {\path{doi:10.1051/0004-6361/201833887}}.

\bibitem{Rasanen:2017ivk}
S.~Rasanen, P.~Wahlman, {Higgs inflation with loop corrections in the Palatini
  formulation}, JCAP 11 (2017) 047.
\newblock \href {http://arxiv.org/abs/1709.07853} {\path{arXiv:1709.07853}},
  \href {https://doi.org/10.1088/1475-7516/2017/11/047}
  {\path{doi:10.1088/1475-7516/2017/11/047}}.

\bibitem{Enckell:2018kkc}
V.-M. Enckell, K.~Enqvist, S.~Rasanen, E.~Tomberg, {Higgs inflation at the
  hilltop}, JCAP 06 (2018) 005.
\newblock \href {http://arxiv.org/abs/1802.09299} {\path{arXiv:1802.09299}},
  \href {https://doi.org/10.1088/1475-7516/2018/06/005}
  {\path{doi:10.1088/1475-7516/2018/06/005}}.

\bibitem{Shaposhnikov:2020fdv}
M.~Shaposhnikov, A.~Shkerin, S.~Zell, {Quantum Effects in Palatini Higgs
  Inflation}, JCAP 07 (2020) 064.
\newblock \href {http://arxiv.org/abs/2002.07105} {\path{arXiv:2002.07105}},
  \href {https://doi.org/10.1088/1475-7516/2020/07/064}
  {\path{doi:10.1088/1475-7516/2020/07/064}}.

\bibitem{Burgess:2009ea}
C.~P. Burgess, H.~M. Lee, M.~Trott, {Power-counting and the Validity of the
  Classical Approximation During Inflation}, JHEP 09 (2009) 103.
\newblock \href {http://arxiv.org/abs/0902.4465} {\path{arXiv:0902.4465}},
  \href {https://doi.org/10.1088/1126-6708/2009/09/103}
  {\path{doi:10.1088/1126-6708/2009/09/103}}.

\bibitem{Barbon:2009ya}
J.~L.~F. Barbon, J.~R. Espinosa, {On the Naturalness of Higgs Inflation}, Phys.
  Rev. D 79 (2009) 081302.
\newblock \href {http://arxiv.org/abs/0903.0355} {\path{arXiv:0903.0355}},
  \href {https://doi.org/10.1103/PhysRevD.79.081302}
  {\path{doi:10.1103/PhysRevD.79.081302}}.

\bibitem{Burgess:2010zq}
C.~P. Burgess, H.~M. Lee, M.~Trott, {Comment on Higgs Inflation and
  Naturalness}, JHEP 07 (2010) 007.
\newblock \href {http://arxiv.org/abs/1002.2730} {\path{arXiv:1002.2730}},
  \href {https://doi.org/10.1007/JHEP07(2010)007}
  {\path{doi:10.1007/JHEP07(2010)007}}.

\bibitem{Hertzberg:2010dc}
M.~P. Hertzberg, {On Inflation with Non-minimal Coupling}, JHEP 11 (2010) 023.
\newblock \href {http://arxiv.org/abs/1002.2995} {\path{arXiv:1002.2995}},
  \href {https://doi.org/10.1007/JHEP11(2010)023}
  {\path{doi:10.1007/JHEP11(2010)023}}.

\bibitem{Lerner:2011it}
R.~N. Lerner, J.~McDonald, {Unitarity-Violation in Generalized Higgs Inflation
  Models}, JCAP 11 (2012) 019.
\newblock \href {http://arxiv.org/abs/1112.0954} {\path{arXiv:1112.0954}},
  \href {https://doi.org/10.1088/1475-7516/2012/11/019}
  {\path{doi:10.1088/1475-7516/2012/11/019}}.

\bibitem{Bezrukov:2013fka}
F.~Bezrukov, {The Higgs field as an inflaton}, Class. Quant. Grav. 30 (2013)
  214001.
\newblock \href {http://arxiv.org/abs/1307.0708} {\path{arXiv:1307.0708}},
  \href {https://doi.org/10.1088/0264-9381/30/21/214001}
  {\path{doi:10.1088/0264-9381/30/21/214001}}.

\bibitem{Kehagias:2013mya}
A.~Kehagias, A.~Moradinezhad~Dizgah, A.~Riotto, {Remarks on the Starobinsky
  model of inflation and its descendants}, Phys. Rev. D 89~(4) (2014) 043527.
\newblock \href {http://arxiv.org/abs/1312.1155} {\path{arXiv:1312.1155}},
  \href {https://doi.org/10.1103/PhysRevD.89.043527}
  {\path{doi:10.1103/PhysRevD.89.043527}}.

\bibitem{Ito:2021ssc}
A.~Ito, W.~Khater, S.~Rasanen, {Tree-level unitarity in Higgs inflation in the
  metric and the Palatini formulation}, JHEP 06 (2022) 164.
\newblock \href {http://arxiv.org/abs/2111.05621} {\path{arXiv:2111.05621}},
  \href {https://doi.org/10.1007/JHEP06(2022)164}
  {\path{doi:10.1007/JHEP06(2022)164}}.

\bibitem{Antoniadis:2021axu}
I.~Antoniadis, A.~Guillen, K.~Tamvakis, {Ultraviolet behaviour of Higgs
  inflation models}, JHEP 08 (2021) 018, [Addendum: JHEP 05, 074 (2022)].
\newblock \href {http://arxiv.org/abs/2106.09390} {\path{arXiv:2106.09390}},
  \href {https://doi.org/10.1007/JHEP05(2022)074}
  {\path{doi:10.1007/JHEP05(2022)074}}.

\bibitem{Giudice:2010ka}
G.~F. Giudice, H.~M. Lee, {Unitarizing Higgs Inflation}, Phys. Lett. B 694
  (2011) 294--300.
\newblock \href {http://arxiv.org/abs/1010.1417} {\path{arXiv:1010.1417}},
  \href {https://doi.org/10.1016/j.physletb.2010.10.035}
  {\path{doi:10.1016/j.physletb.2010.10.035}}.

\bibitem{Ema:2020zvg}
Y.~Ema, K.~Mukaida, J.~van~de Vis, {Higgs inflation as nonlinear sigma model
  and scalaron as its $\sigma$-meson}, JHEP 11 (2020) 011.
\newblock \href {http://arxiv.org/abs/2002.11739} {\path{arXiv:2002.11739}},
  \href {https://doi.org/10.1007/JHEP11(2020)011}
  {\path{doi:10.1007/JHEP11(2020)011}}.

\bibitem{Mikura:2021clt}
Y.~Mikura, Y.~Tada, {On UV-completion of Palatini-Higgs inflation}, JCAP
  05~(05) (2022) 035.
\newblock \href {http://arxiv.org/abs/2110.03925} {\path{arXiv:2110.03925}},
  \href {https://doi.org/10.1088/1475-7516/2022/05/035}
  {\path{doi:10.1088/1475-7516/2022/05/035}}.

\bibitem{Sfakianakis:2018lzf}
E.~I. Sfakianakis, J.~van~de Vis, {Preheating after Higgs Inflation:
  Self-Resonance and Gauge boson production}, Phys. Rev. D 99~(8) (2019)
  083519.
\newblock \href {http://arxiv.org/abs/1810.01304} {\path{arXiv:1810.01304}},
  \href {https://doi.org/10.1103/PhysRevD.99.083519}
  {\path{doi:10.1103/PhysRevD.99.083519}}.

\bibitem{Bezrukov:2008ut}
F.~Bezrukov, D.~Gorbunov, M.~Shaposhnikov, {On initial conditions for the Hot
  Big Bang}, JCAP 06 (2009) 029.
\newblock \href {http://arxiv.org/abs/0812.3622} {\path{arXiv:0812.3622}},
  \href {https://doi.org/10.1088/1475-7516/2009/06/029}
  {\path{doi:10.1088/1475-7516/2009/06/029}}.

\bibitem{Garcia-Bellido:2008ycs}
J.~Garcia-Bellido, D.~G. Figueroa, J.~Rubio, {Preheating in the Standard Model
  with the Higgs-Inflaton coupled to gravity}, Phys. Rev. D 79 (2009) 063531.
\newblock \href {http://arxiv.org/abs/0812.4624} {\path{arXiv:0812.4624}},
  \href {https://doi.org/10.1103/PhysRevD.79.063531}
  {\path{doi:10.1103/PhysRevD.79.063531}}.

\bibitem{Ema:2016dny}
Y.~Ema, R.~Jinno, K.~Mukaida, K.~Nakayama, {Violent Preheating in Inflation
  with Nonminimal Coupling}, JCAP 02 (2017) 045.
\newblock \href {http://arxiv.org/abs/1609.05209} {\path{arXiv:1609.05209}},
  \href {https://doi.org/10.1088/1475-7516/2017/02/045}
  {\path{doi:10.1088/1475-7516/2017/02/045}}.

\bibitem{Bauer:2010jg}
F.~Bauer, D.~A. Demir, {Higgs-Palatini Inflation and Unitarity}, Phys. Lett. B
  698 (2011) 425--429.
\newblock \href {http://arxiv.org/abs/1012.2900} {\path{arXiv:1012.2900}},
  \href {https://doi.org/10.1016/j.physletb.2011.03.042}
  {\path{doi:10.1016/j.physletb.2011.03.042}}.

\bibitem{Rubio:2019ypq}
J.~Rubio, E.~S. Tomberg, {Preheating in Palatini Higgs inflation}, JCAP 04
  (2019) 021.
\newblock \href {http://arxiv.org/abs/1902.10148} {\path{arXiv:1902.10148}},
  \href {https://doi.org/10.1088/1475-7516/2019/04/021}
  {\path{doi:10.1088/1475-7516/2019/04/021}}.

\end{thebibliography}
\end{document}